\documentclass[ALICE,manyauthors]{cernphprep}

\usepackage[comma,square,numbers,sort&compress]{natbib}
\usepackage{hyperref}
\usepackage{color}
\usepackage[dvipsnames]{xcolor}
\usepackage{lineno}
\usepackage{booktabs}
\usepackage{xspace}
\usepackage[littleEndian,numberBitsAbove,numberFieldsTwice]{bitpattern}

\usepackage{siunitx}

\usepackage[T1]{fontenc}

\DeclareGraphicsExtensions{.pdf} 
\graphicspath{{./figs/}}

\newcommand{\pt}{\ensuremath{p_{\mathrm T}}\xspace}

\newcommand{\sqrts}[1]   {\mbox{$\sqrt{s} = #1$ TeV}\xspace}

\newcommand{\zvtx}{\ensuremath{z_{\mathrm{vtx}}}\xspace}

\newcommand\gevc{GeV/$c$\xspace}

\newcommand{\K}{\ensuremath{\mathrm{K_S^0}}\xspace}
\newcommand{\lam}{\ensuremath{(\Lambda+\overline{\Lambda})}\xspace}
\newcommand{ \dphi}{\ensuremath{\Delta\varphi}\xspace}
\newcommand{ \deta}{\ensuremath{\Delta\eta}\xspace}
\newcommand{ \pttrig}{\ensuremath{p_\mathrm{T}^\mathrm{trigg}}\xspace}
\newcommand{ \ptassoc}{\ensuremath{p_\mathrm{T}^\mathrm{assoc}}\xspace}
\newcommand{\vzero }{\ensuremath{\mathrm{V^{0}}}\xspace}

\begin{document}%

\begin{titlepage}
\PHyear{2021} %
\PHnumber{146}      
\PHdate{21 July}  
%

\title{\boldmath$\K$- and (anti-)\boldmath$\Lambda$-hadron correlations in pp collisions at $\mathbf{\sqrt{ \emph s}}$ = 13 TeV}
\ShortTitle{K$^{0}_{\rm S}$- and (anti-)$\Lambda$-hadron correlations in pp collisions at $\sqrt{s} = 13$ TeV}   

\Collaboration{ALICE Collaboration\thanks{See Appendix~\ref{app:collab} for the list of collaboration members}}
\ShortAuthor{ALICE Collaboration} 

\begin{abstract}
 Two-particle azimuthal correlations are measured with the ALICE apparatus in pp collisions at \sqrts{13} to explore strangeness- and multiplicity-related effects in the fragmentation of jets and the transition regime between bulk and hard production, probed with the condition that a strange meson (\K) or baryon ($\Lambda$) with transverse momentum $\pt>3$ \gevc is produced.
Azimuthal correlations between kaons or $\Lambda$ hyperons with other hadrons are presented at midrapidity for a broad range of the trigger ($3 < \pttrig < 20$ \gevc) and associated particle \pt (1 \gevc $ < \ptassoc < \pttrig$), for minimum-bias events and as a function of the event multiplicity.
The near- and away-side peak yields are compared for the case of either \K or $\Lambda$($\overline{\Lambda}$) being the trigger particle with that of inclusive hadrons (a sample dominated by pions). In addition, the measurements are compared with predictions from PYTHIA 8 and EPOS LHC event generators.
\end{abstract}

\end{titlepage}
\setcounter{page}{2}

%
%
\section{Introduction}
\label{sec:intro}

Particle production as a function of the event charged-particle multiplicity in proton-proton (pp) collisions at the LHC has revealed interesting patterns.
Clearly, in the soft (bulk) particle production domain with low transverse momentum ($\pt\lesssim\ $4 \gevc), several experimental measurements indicate features in high-multiplicity pp collisions similar to those observed in nucleus--nucleus collisions. These include long-range correlations in pseudorapidity~\cite{Khachatryan:2010gv,Khachatryan:2015lva},
large azimuthal anisotropies~\cite{Aad:2015gqa,Khachatryan:2016txc} 
and strangeness production~\cite{ALICE:2017jyt,Acharya:2019kyh}.

These measurements are theoretically interpreted in terms of a combination of initial-state collective dynamics (colour-glass condensate)~\cite{Dusling:2012cg} 
or as a hydrodynamic-like (final-state) collective flow~\cite{Schenke:2020mbo}. Quantifying the relative contributions of initial- and final-state phenomena is a challenge, both experimentally and theoretically (see reviews~\cite{Dusling:2015gta,Schenke:2021mxx}). These phenomena are also modelled in Monte Carlo (MC) event generators, like PYTHIA8~\cite{Sjostrand:2014zea} or EPOS LHC~\cite{Pierog:2013ria}. 
For example, a basic experimental finding in pp collisions, the increase of the average transverse momentum with the event multiplicity is realized in these two models very differently.
In PYTHIA 8, in events with several partonic scatterings, termed Multiple Parton Interactions (MPI), the respective color strings cut (reconnect) each other, leading to a~redistribution of energy from particle production to transverse momentum.
In the EPOS LHC model, a~parametrised hydrodynamic evolution of a~small volume with high density of thermalised matter (core) is used.
In both models the respective parameters were tuned using the Run 1 data at the LHC, without explicit inclusion of particle correlations~\cite{Skands:2014pea,Pierog:2013ria}. 

In PYTHIA8, 
correlations among the final state hadrons are realized through transversely extended strings, exerting on each other transverse shoves~\cite{Bierlich:2017vhg} that mimic  collective dynamics, akin to that of a (long-lived) quark--gluon medium.
The shoving model of hadronisation was recently used to discern within a PYTHIA8 study~\cite{Bierlich:2019ixq} possible effects of jet quenching in pp collisions. 
This prominent characteristic of nucleus--nucleus collisions, remains undetected in high-multiplicity pp or p--Pb collisions~\cite{Adam_2015}, perhaps not surprisingly, given the much smaller spatial extension of the dense system, compared to the~nucleus--nucleus case. 
Experimentally, it was recently shown  that in pp collisions the near-side long-range (in pseudorapidity) ridge yield (of bulk particles) in high-multiplicity events remains present for events which are additionally biased, through either a $Z$ boson~\cite{Aaboud:2019mcw}, a leading high-\pt particle or a jet~\cite{Acharya:2021gnt}. 
These findings are interesting per se and also motivate the~quest to find or exclude jet quenching in high-multiplicity events in small collision systems, with differences in the observed effects on gluon, light and heavy quark jets.

At LEP, differences between quark- and gluon-initiated jets in e$^+$-e$^-$ annihilations have been revealed in several measurements. Gluon jets are characterised by a larger charged-particle multiplicity than quark jets~\cite{Abbiendi:2002,delphi_quarkGluon}. Moreover, in the~relative production of \K mesons and $\Lambda$ hyperons to charged particles, it was found that the relative production of $\Lambda$ is $\approx$30\% higher in gluon than in quark jets, while the relative \K production was found to be approximately the same~\cite{Ackerstaff:1999}.  

In the present article, such studies of particle production and correlations are continued, exploring the~effect of a stran\-ge\-ness bias, both in form of a meson (\K) or a baryon ($\Lambda$) with $\pt>3$ \gevc.
The~correlations between kaons or $\Lambda$($\overline{\Lambda}$) hyperons with other hadrons are studied at midrapidity in pp collisions at \sqrts{13} for a~broad range of the trigger ($3 < \pttrig < 20$ \gevc) and associated particle \pt (1 \gevc\ $< \ptassoc < \pttrig$), for minimum bias events and as a function of the event multiplicity measured at forward rapidities.
Such correlations encode effects of fragmentation, hadronisation and parton showering as well as possible collectivity and jet quenching. 
The complex overlap of these aspects is probed through the experimental handles of particle species and \pt, inducing different kinematic and flavour biases.  
The near- and away-side peak yields are compared for the case of either \K or (anti-)$\Lambda$ as a trigger particle with that of inclusive hadrons (a sample dominated by pions). The measurements are, in addition, compared with the~PYTHIA 8 and EPOS LHC event generators.

The article is organised as follows: Section~\ref{sec:exp} outlines the experimental setup and the data sample; Section~\ref{sec:ana} describes the analysis, while Section~\ref{sec:res} presents the results; a summary and an outlook are given in Section~\ref{sec:conc}.

\section{Experiment and data sample}
\label{sec:exp}
The inclusive charged hadron, \K meson and (anti-)$\Lambda$ hyperon identification at midrapidity is performed using the tracking detectors of the ALICE central barrel located in a solenoidal magnet, which provides a magnetic field of 0.5 T oriented along the beam direction. A detailed description of the ALICE experiment and its performance can be found in~\cite{Abelev:2008aa, Abelev:2014ffa}. 

\subsection{Event selection}
For the data taking, a minimum bias (MB) trigger is employed,  provided by the V0 detector, which consists of two forward scintillator arrays covering the pseudorapidity ranges $-3.7<\eta<-1.7$ and $2.8~<~\eta~<~5.1$. The MB trigger signal consists of a coincident signal in both arrays.
All events selected in this analysis are required to have a reconstructed primary collision vertex (PV) within the~longitudinal interval $|\zvtx|<10$~cm from the nominal interaction point in order to ensure uniform detector performance. Beam-gas events are rejected using timing cuts with the~V0 detector. Moreover pile-up events are rejected based on the Silicon Pixel Detector (SPD) information. 
The total number of analysed pp collisions at \sqrts{13}, measured during the LHC Run~2 data-taking period in years 2016--2018 by ALICE is $1.58 \times 10^9$ corresponding to integrated luminosity of about 27 nb$^{-1}$.

\subsection{Multiplicity selection}

The correlation functions are calculated for six event classes (0--1\%, 1--3\%, 3--7\%, 7--15\%, 15--50\%, 50--100\%), selected on the event activity via the multiplicity in the forward and backward direction, measured with the V0 detector within the acceptance described above. The events are selected based on percentiles of the summed signal in the two V0 detectors (V0M), for instance the 0--1\% and 50--100\% classes are the events with the highest and lowest range of the V0M signal. The trigger efficiency is not accounted for in the above multiplicity ranges. The intervals corrected for the trigger efficiency are: respectively; 0--0.92\%, 0.92--2.74\%, 2.74--6.40\%, 6.40--13.44\%, 13.44--46.12\%, 46.12--100\%~\cite{alicecollaboration2020pseudorapidity}.

For the MC event generators, the multiplicity classes are  selected with the trigger-corrected percentile calculation, applied to the distribution of charged primary particles produced in the $\eta$ acceptance of the~V0 detectors.

\subsection{Primary hadron and \vzero selection}
Primary charged tracks (denoted as h) are reconstructed in the pseudorapidity range $|\eta|<0.8$ using the~Inner Tracking System (ITS), which consists of six layers of silicon detectors around the beam pipe, and the Time Projection Chamber (TPC), consisting of a large cylindrical drift volume filled with nearly \SI{90}{m^3} either of Ar/CO\textsubscript{2} (88/12, in 2016 and 2018) or Ne/CO\textsubscript{2}/N\textsubscript{2} (90/10/5 in 2017)  and read out by multi-wire proportional chambers. 
Combining the information from these two detectors, the primary charged-track sample is created by applying selection criteria in order to suppress the contamination from secondary particles following previous studies~\cite{ALICE:2018vuu}. 
The number of crossed pad rows in the TPC is required to be at least 70 (out of a maximum of 159) and the minimal ratio to the number of findable clusters (geometrically possible assignable clusters to a track) is 0.8. 
Only tracks with a fraction of shared clusters with other tracks smaller than 0.4 are accepted.  
The~distance of closest approach (DCA) to the PV is required to be within an elipsoid with semi-axes of 2.4 cm and 3.2 cm in the~$xy$-plane and $z$-direction, respectively. 
Every track is required to have a fit quality for both TPC and ITS, characterised by goodness-of-fit values $\chi^2$ per cluster smaller than 4 and 36 for the TPC and ITS, respectively. 
Only tracks with a hit in the two most inner layers of ITS are selected. The kink topologies produced by decays are rejected.  The selected sample of primary charged particles is dominated by hadrons. The electrons constitute less than 1\%.

 \begin{table}[h!]
 \centering
\caption{ Selection criteria for $\rm V^0$ candidates based on the topological variables. } \label{tab:topolVar}
 \begin{tabular}{lcc}
\toprule
Selection criterion & $\rm K_S^0$  & $\Lambda(\overline{\Lambda})$  \\ \hline
 Absolute value of rapidity & $<0.5$          & $<0.5$  \\
 Decay radius (cm)       &$>0.5$&$>0.5$\\
 ${\rm DCA}_{xy}$ of daughter track to PV (cm)       &$>0.06$&$>0.06$\\
 ${\rm DCA}_{xy}$ between daughter tracks ($n_\sigma$) & $<1$ & $<1$ \\
 cos($\theta_{\mathrm{PA}}$) & $>0.97$ & $>0.995$ \\
 Proper lifetime (cm) & $<20$& $<30$ \\
 Competing rejection (GeV/$c^2$) & $>0.005$& $>0.01$ \\
  Invariant mass (GeV/$c^2$) & $m_\K\pm3\sigma$& $m_{\Lambda(\overline{\Lambda})}\pm3\sigma$ \\
\bottomrule
\end{tabular}
 \end{table}

The $\rm K_S^0$ mesons and $\Lambda$($\overline{\Lambda}$) baryons (\vzero particles) are reconstructed in the rapidity range $|y|<0.5$ via their most probable decay channels~\cite{Tanabashi:2018} and exploiting their characteristic (\vzero) decay topology:

\begin{equation*}
\K\rightarrow\pi^{+}+\pi^{-} (69.2\%),
\end{equation*}
\begin{equation*}
\Lambda\rightarrow \mathrm{p}+\pi^{-} (63.9\%),
\end{equation*}
\begin{equation*}
\overline{\Lambda}\rightarrow \mathrm{\overline{p}}+\pi^{+} (63.9\%).
\end{equation*}

The identification and reconstruction follows previous measurements presented in~\cite{Acharya:2020} and~\cite{Aamodt1:2010}. The~identification of the daughter tracks 
is performed via specific energy loss d$E$/d$x$ in the TPC, which is required to be within $\pm$3$\sigma$ from the expected mean value for pions or protons (for more details see~\cite{Abelev:2014ffa}). The~track quality criteria are the same as for the primary global tracks described above.
The pairs of identified daughter tracks are combined to \vzero candidates, which are accepted if their invariant mass is within 3$\sigma$ (1$\sigma$ for \K is in the range 0.0039-0.0075 GeV/$c^2$ and for $\Lambda$ in 0.0021-0.0033 GeV/$c^2$ depending on \pt) from the nominal value. The combinatorial contribution is suppressed by applying  selection criteria based on the topological variables summarised in Table~\ref{tab:topolVar}. 
Here, the \vzero decay radius is the distance between the point where the \vzero decays (secondary vertex) and the~PV. The~${\rm DCA}_{xy}$ to PV is the distance of closest approach between the daughter track and the PV. The $\theta_{\mathrm{PA}}$ refers to the~pointing angle, which is the angle between the momentum vector of the \vzero candidate and the line connecting the primary and secondary vertex. The reconstructed proper lifetime of an individual particle is defined as $ mL/p$, where $m$ is the particle mass, $L$ is the distance between primary and secondary vertex and $p$ is the particle momentum. The mean life $c\tau$ is listed in~\cite{Tanabashi:2018} and its value is 2.68 cm and 7.89 cm for \K and $\Lambda(\overline{\Lambda})$, respectively. It can happen that a certain pair reconstructed as $\rm K_S^0$ candidate can have an invariant mass of $\Lambda(\overline{\Lambda})$ under the p$\pi$ assumption for the daughter tracks. Such pairs are not accepted neither as \K nor as $\Lambda(\overline{\Lambda})$ candidates. Table~\ref{tab:topolVar} quotes under the competing rejection entry the mass on which this criterion is applied for both \K and $\Lambda$. %
Besides the topological selections, a bunch-off pile-up (by the high frequency collisions, some tracks from previous bunch-crossing remain in TPC when current collision happen) removal criterion is required where at least one of the \vzero daughter tracks is reconstructed both in ITS and TPC or has a  signal in the Time-Of-Flight detector. 

\begin{figure}[hbt]
    \centering
    \includegraphics[width=0.95\textwidth]{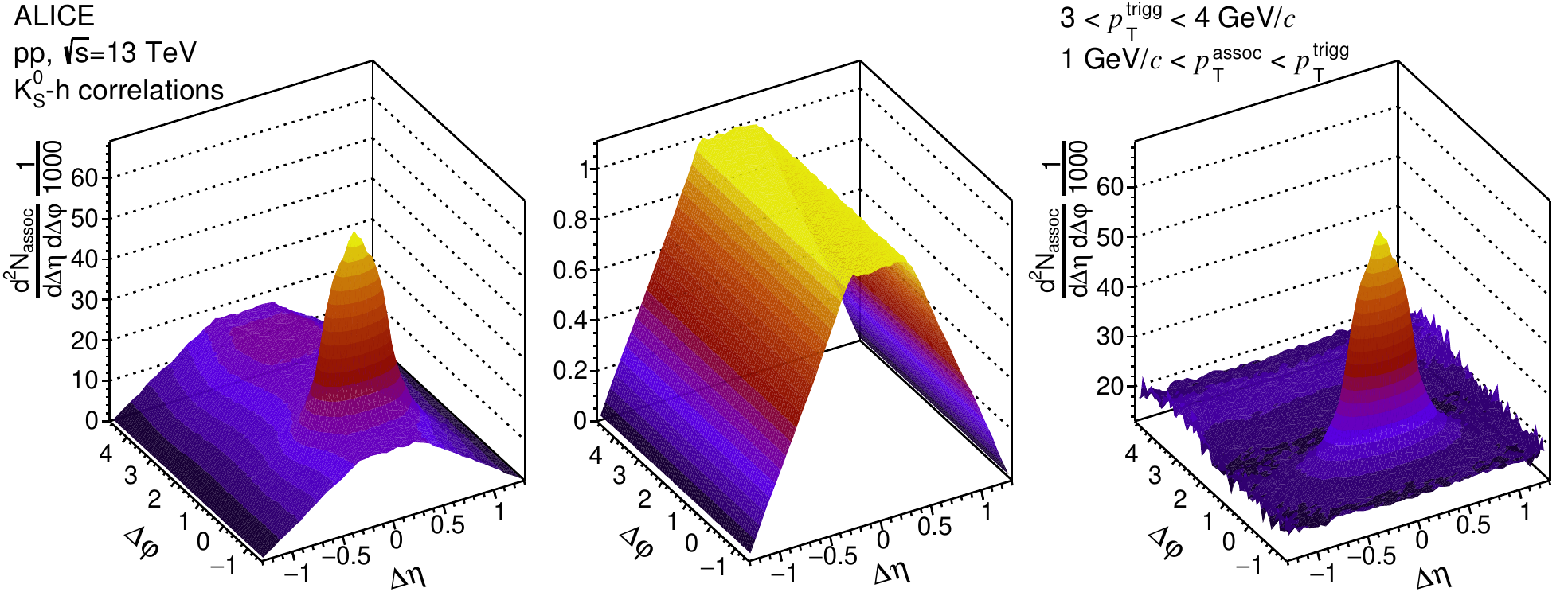}
    \caption{An example of the raw same-event (left), mixed-event (middle) and final (mixed-event scaled, right) two-dimensional correlation function for \K-hadrons. The correlation functions were scaled with 1/1000 for better visibility. The plateau in the left and middle plot is caused by non-equal selection in $\eta$ of the trigger and associated particle.}
    \label{fig:mixing}
\end{figure}

\section{Analysis}
\label{sec:ana}
\subsection{The correlation function}

In the dihadron correlation approach, 
the first hadron is the 
trigger particle, here either a primary charged particle (hadron) or an identified \K meson or $\Lambda$($\overline{\Lambda}$) hyperon with \pt in range 3-20 \gevc. Since the $\Lambda$-h and $\overline{\Lambda}$-h correlation functions are compatible, as expected for this collision energy, the results are combined and reported in the following as \lam-h.
The second particle is the 
associated particle, in this case, always a primary charged particle with a kinematic requirement 1 \gevc $< $\ptassoc $< $ \pttrig. 
By calculating the differences in the azimuthal angle and pseudorapidity for each of such pairs, three types of correlation functions are constructed: h-h, \K-h and \lam-h. 
For h-h correlations, pairs with invariant mass (IM) within $\pm$5 MeV/$c^2$ of the mass of \K or $\Lambda$($\overline{\Lambda}$) or from $\gamma$ conversions are not accepted. The contribution from decays of the K$^*$(892) and $\phi$ mesons,  $\Delta$ resonances and D mesons was checked and found negligible.
In the case of \lam-h correlations, this restriction is applied to pairs with IM of a cascade ($\Sigma$, $\Xi$, $\Omega$).An example of a raw \K-h correlation function is shown in Fig.~\ref{fig:mixing} (left panel). At $(\dphi,\deta) = (0,0)$, one can observe the~near-side peak, which originates mostly from particle pairs fragmented within the~same jet. Bose-Einstein correlations, strong decays of high-mass resonances and final state interactions may have also a small contribution for the h-h case. Due to momentum conservation, jets are produced back-to-back in the transverse plane. Thus, a second peak around $\pi$ in \dphi is expected, which is smeared in the \deta direction, because the particles can obtain an additional longitudinal boost related to the varied center-of-mass frame of the partonic collision. In the~selection of the~trigger particle, the near-side jet is fully reconstructed in the longitudinal direction, but the away-side jet is not necessarily (fully) within the detector acceptance. 
The procedure of getting fully corrected 2-dimensional per-trigger yield is schematically written in Eq~\ref{eqcorr}. Here, $\frac{{\rm d}^2N_{\mathrm{pair}}^{\rm raw}}{{\rm d}\Delta\varphi {\rm d} \Delta\eta}(\Delta\varphi,\Delta\eta)$ is the uncorrected correlation function, $\varepsilon_{\mathrm{trigg}}$,$\varepsilon_{\mathrm{assoc}}$ and $\varepsilon_{\mathrm{pair}}$ are correction factors further described in Sec.~\ref{sec:corr} and $N_{\mathrm{trigg}}$ is the number of the trigger particles. Afterwards, the 2-dimensional per-trigger yield is projected on the $\Delta\varphi$ axis and integrated (see Eq.~\ref{eqyield}) in the intervals $|\dphi| <0.9$ and $ |\dphi -\pi| < 1.4$ to obtain the near-side and away-side yield, respectively, denoted as $Y_{\Delta\varphi}$ in Eq.~\ref{eqyield}. 

\begin{equation}
    \frac{{\rm d}^2 N_ {\mathrm{pair}}}{{\rm d}\Delta\varphi {\rm d}\Delta\eta}(\Delta\varphi,\Delta\eta) =\frac{1}{N_{\mathrm{trigg}}}  \frac{1}{\varepsilon_{\mathrm{trigg}}}\frac{1}{\varepsilon_{\mathrm{assoc}}} \frac{{\rm d}^2N_{\mathrm{pair}}^{\rm raw}}{{\rm d}\Delta\varphi {\rm d} \Delta\eta}(\Delta\varphi,\Delta\eta)\frac{1}{\varepsilon_{\mathrm{pair}}}
 \label{eqcorr}
\end{equation}
 \begin{equation}
 Y_{\Delta\varphi} = \int_{\Delta\varphi_1}^{\Delta\varphi_2} \frac{{\rm d}N}{{\rm d}\Delta\varphi}{\rm d}\Delta\varphi 
  \label{eqyield}
 \end{equation}

\subsection{Corrections}
\label{sec:corr}
The corrections are described in the same order as they were applied to the data. 

All MC-based corrections are calculated using events from PYTHIA8.210 (Monash 2013 tune)~\cite{Sjostrand:2007gs,Skands:2014pea}, with particle propagation through the
detector by means of GEANT3~\cite{A32_RefGeant3}.
The detection inefficiencies are corrected with the~single particle efficiency factor, calculated in MC and applied as weight ($1/\varepsilon_{\mathrm {trigg}}\times 1/\varepsilon_{\mathrm {assoc}}$) for each pair. This factor was calculated separately for trigger and associated particles as a function of \pt, $\eta$, $\varphi$ and PV position. In the case of primary charged particles, a \pt-dependent contamination factor is also part of the weight to account for the amount of secondary particles in the~sample. This is defined as a ratio of only primary tracks to all reconstructed ones.

Imperfect detector acceptance within $|\eta|\ <\ $
0.8 range is corrected with the mixed-event method, where trigger particles from one event are correlated with associated particles from different events. Thus, no physical correlations are present. The mixed-event correlation function has a typical triangular shape determined by the $\eta$ acceptance. An example of this function is shown in the middle plot of Fig.~\ref{fig:mixing} where a plateau is visible. This is caused by different ranges in $\eta$ for trigger (\K) and associated particles (h). The mixed-event correlation is already scaled to unity with a scaling factor equal to the average of bins with \deta = 0. In the following, the actual correlation function is divided by the mixed-event one to eliminate the detector acceptance effects as illustrated in Fig.~\ref{fig:mixing}. This correction is schematically written as $1/\varepsilon_{\mathrm{pair}}$ in Eq.~\ref{eqcorr}. In some cases, due to the finite binning in multiplicity and PV position in z-direction, the mixed-event correlation does not match the~shape of the~background perfectly. For this reason, a~so called ``wing`` correction is performed. Here the~correlation function is scaled once more with a 2D distribution constant in \dphi and dependent on \deta in order to get a flat distribution in \deta at the away-side. This correction is never larger than 2\% and only affects the h-h correlation function. A similar effect was observed also in a previous analysis~\cite{Adam:2016ckp}.

For the reconstruction of \K mesons and $\Lambda$($\overline{\Lambda}$) baryons, some of the candidates selected with the topological criteria are in fact combinatorial  background. 
Since the~shape of the~correlation function does not need to be the same for the signal and background, a second correlation function is calculated, where candidates from two intervals from outside the invariant mass peak ($m_\vzero - 9\sigma$ to $m_\vzero-6\sigma$ and $m_\vzero+6\sigma$ to $m_\vzero+9\sigma$) are taken as trigger particles. These give the same width as the signal region in the invariant mass spectrum. The second ``side-band`` correlation function is subtracted from the signal one. The~number of trigger particles is in addition corrected for purity, defined as a ratio of number of signal \vzero candidates over all candidates within the invariant mass acceptance region.   

In the case of $\Lambda$($\overline{\Lambda}$) being the trigger particle, the feed-down contribution from decays of $\Xi$ baryons (reconstructed following~\cite{Acharya:2019kyh}) baryons is subtracted in a similar way as for the combinatorial background. For this case, the $(\Xi^-+\overline{\Xi}^+)$-h correlation function in every \pt and multiplicity bin is calculated, scaled with the detection efficiency of $\Lambda$($\overline{\Lambda}$) from $\Xi$ decays and subtracted from the \lam-h correlation function. Similarly, the feed-down fraction is subtracted from the number of trigger particles. It is assumed that the production rates of charged and neutral $\Xi$ baryons are equal and the feed-down fraction from $\Omega$ is negligible. This correction has an effect of 5\% on the final near-side yields for low \pt and smaller than 1\% for high \pt. 

After projecting the per-trigger yield on the \dphi axis, the underlying event background is subtracted with the ZYAM (Zero Yield At Minimum) method~\cite{Trainor:2010dlo}. The background is assumed to be flat and estimated as the average value of six bins outside the jet peaks to reduce the statistical fluctuations.

\subsection{Systematic uncertainties}

 \begin{table}[h!]
 \centering
\caption{ Summary of the main sources and values of the relative systematic uncertainties (expressed in \%) for the~per-trigger yields in the MB sample. The abbreviation ``negl.`` stands for negligible (smaller than 0.1\%) and ``rej.`` means that this variation was rejected due to the Barlow criterion.} \label{tab:syst}
 \begin{tabular}{lcccccc}
\toprule
& \multicolumn{2}{c}{h-h} & \multicolumn{2}{c}{$\rm K_S^0$-h}  & \multicolumn{2}{c}{$\Lambda(\overline{\Lambda})$-h}  \\ \hline
& near & away & near & away & near & away \\ \hline
 \dphi window & 0.3 & 0.4 & 0.5 & 0.7 & 0.7 & 0.5  \\
 PV along the z-axis (\zvtx)      & negl. & negl. & 1.2 & 1.7 & 0.6 & 0.7  \\
 Binning in \zvtx       & negl. & 0.4 & 0.8 & 1.6 & 0.5 & 1.2  \\
 Yield calculation  & 1.0 & negl. & 1.1 & 0.7 & 0.3 & 0.4   \\
 Pedestal subtraction  & 0.8 & 1.9 & 0.4 & 1.6 & 1.0 & 2.0  \\
 $\Delta\eta$ range & 0.5 & -- & 1.2 & -- & 1.0 & --  \\
 Mixing scale & negl. & negl. & 0.7 & 0.9 & 0.3 & 0.5  \\
 Topological variables & -- & -- & 1.5 & 3.5 & 3.0 & 3.1  \\
 Invariant mass range & -- & -- &  rej. &  rej. & 1.2 & 1.8  \\
 Primary track selection & 0.3 & 0.7 & 1.1 & 1.8 & 2.2 & 0.4   \\
 Wing correction & 1.2 & 1.8 & -- & -- & 0.7 & 0.8  \\
 $\Xi$ topological variables & -- & -- & -- & -- & 0.4 & 1.6  \\
 MC closure & negl. & negl. & negl. & negl. & 2.5 & negl.   \\ \hline
 Total & 1.9 & 2.7 & 3.1 & 5.1 & 5.1 & 4.8 \\
\bottomrule
\end{tabular}
 \end{table}

The sources of systematic uncertainties of the per-trigger yields in the minimum bias sample are listed in Tab.~\ref{tab:syst}. These are estimated by varying track-selection criteria and other parameters in the analysis.
The~significance of each source of systematic uncertainty was checked according to the Barlow criterion~\cite{Barlow:2002llk}. Within this procedure a threshold value (1 $\sigma$) is set, based on which each variation can be checked, whether it is within statistical fluctuations or a real systematic difference. If a certain variation did not pass the test, this contribution was not accounted for in the total systematic uncertainty, which was calculated as a quadrature sum of the individual contributions. For the ratios of yields, the systematic uncertainties are calculated separately which causes cancellation of correlated uncertainties.

For the uncertainty related to the \dphi integration window, the window is varied around the nominal values ($|\dphi|<~0.9$ and $|\dphi-\pi|<1.4$) within $\pm$0.1. For the~yields for the h-h correlations, on both near- and away-side, the contribution to the total uncertainty is around 0.4\% for all multiplicity classes. For the~yields for \K-h and \lam-h correlations, the value varies within 0.4--2\% for both near- and away-side.  

The PV selection along the $z$-axis (\zvtx) is decreased from $\pm 10$ cm to $\pm 7$ cm from the interaction point in order to estimate the uncertainty connected to the detector acceptance effects. The uncertainty is smaller than 0.3\% in all multiplicity classes for the yields from h-h correlation function. It is in the range 0.7--2.3\% and 0.7--2.7\% for the near-side yield in case of \K-h and \lam-h yields, respectively. For the~away-side, this source contributes with 1.7--4.5\% and 0.7--4.9\% in case of \K-h and \lam-h yields, respectively.

The number of bins in \zvtx used for the event-mixing classes is changed from 9 to 7 to account for the~uncertainty connected with the detector acceptance. For the yields  triggered with an unidentified hadron, the contribution from this source is smaller than 0.5\% at both sides for all multiplicity classes. This uncertainty is in the range 0.5--2.7\% and 0.4--1.5\% for the near-side yield triggered with \K and \lam, respectively and within 1.2--5.2\% and 0.8--2.8\% for the away-side.

The contribution to the systematic uncertainty resulting from the yield calculation method is estimated by fitting each jet peak with a double-gaussian function and integrating the fit function to calculate the~per-trigger yield instead of calculating the yield directly by the bin counting method as default. This leads to an uncertainty around 1\% for the near-side and to a value smaller than 0.2\% for the~away-side for the~h-h yields in all multiplicity classes. For most multiplicity classes this source was rejected by the~Barlow criterion for the \K trigger. The non-rejected contribution is 1.1\% and 0.7\% for the near- and away-side, respectively. The accounted contribution to the uncertainty of yields triggered with \lam is in the~range 0.3--0.8\% and 0.2--3.4\% for the near-side and away-side yields, respectively.  

For the variation of the underlying event subtraction method, which takes the average value of 6 bins from the left and right side of the near-side peak, a constant fit in ranges $[-\pi/2, -1]$ and $[1,\pi/2]$ is used, leading to an estimated uncertainty around 0.6\% (1.5\%), 2\% (2.2\%) and 1.8\% (4.5\%) for the~near- (away-~) side yield from the unidentified hadron-, \K-, and \lam-triggered correlation functions, respectively. 

The $\Delta\eta$ range is varied within 0.1 around its nominal value $|\deta|<1$ in order to estimate the uncertainty related to the near-side jet acceptance. This is estimated to be within 0.3--0.9\%, 0.6--1.9\%, 0.4-2.4\% for h-h, \K-h and \lam-h yields in all multiplicity classes, respectively.  

The scale factor for the mixed-event correlation function is varied, which gives a negligible contribution to the total systematic uncertainty for h-h yields for both sides. This contribution for \K-h (\lam-h) yields is estimated as 0.7--1.5\% (0.2--0.4\%) and 0.9--2\% (0.2--0.5\%) for the near and away-side yields, respectively, in different multiplicity classes. 

In order to estimate the systematic uncertainty connected to the \vzero reconstruction, the values for the~topological selection  are varied around the nominal values. Its value is, for different multiplicity classes, in the range 1.5--5.8\% (1.9--5.6\%) and 2.2--7.5\% (2--5.5\%) for the \K \lam triggered yields at the near- and away-side, respectively. 

The ranges for the signal and for background in the invariant mass distributions are varied in order to estimate the uncertainty related to the~subtraction of the contribution from misidentified \K$\ $ or \lam. This source is rejected by the~Barlow criterion for the \K-h yields and has a value in the range 0.5--3.9\% and 1.1--4.3\% for the~\lam-h triggered yields, for the near- and away-side, respectively.

The systematic uncertainty associated with the primary track selections is estimated by selecting tracks with slightly varied criteria. These are the same as the ones used for global tracks, but there is a~tighter and \pt-dependent DCA requirement in the $xy$-plane, which means that tracks with a DCA in the $xy$-plane larger than $0.0105+0.0350/\pt^{1.1}$ are rejected. This uncertainty is smaller than 0.7\% for both the~near- and away-side yield for h-h for all multiplicity classes. The uncertainty intervals for \K (\lam) triggered yields are estimated as 0.9--2.4\% (0.4--3.5\%) and 1.3--3.9\% (0.2--3.1\%) for the near- and away-side yields, respectively. 

The range used for the estimation of the wing correction scaling factor is varied in order to calculate the~uncertainty related to this method. This contribution is not dependent on the event multiplicity.  

The $\Xi^-(\overline{\Xi}^+)$ reconstruction uncertainty contributes to the uncertainty of yields triggered by \lam. This contribution is estimated by varying the topological selection of $\Xi^-(\bar{\Xi}^+)$ hyperons around their nominal values. This uncertainty is in the range 0.2--4\% (0.2--3.9\%) for the near(away)-side yields for events in all multiplicity classes.

The correction procedure is checked with a Monte Carlo closure test. Two correlation functions are calculated, the first one with generated MC particles and the second one with MC particles reconstructed after GEANT3 propagation using the full reconstruction and correction chain as for the experimental data. The ratio of these two correlation functions is expected to be unity. This is the case for the h-h and \K-h correlation functions, but there is a residual departure from unity for \lam-h correlation function at the near-side of up to 2.5\%, which is accounted as a systematic uncertainty.

\section{Results and discussion}
\label{sec:res}

\begin{figure}[hbt!]
    \centering
     \includegraphics[width=\linewidth]{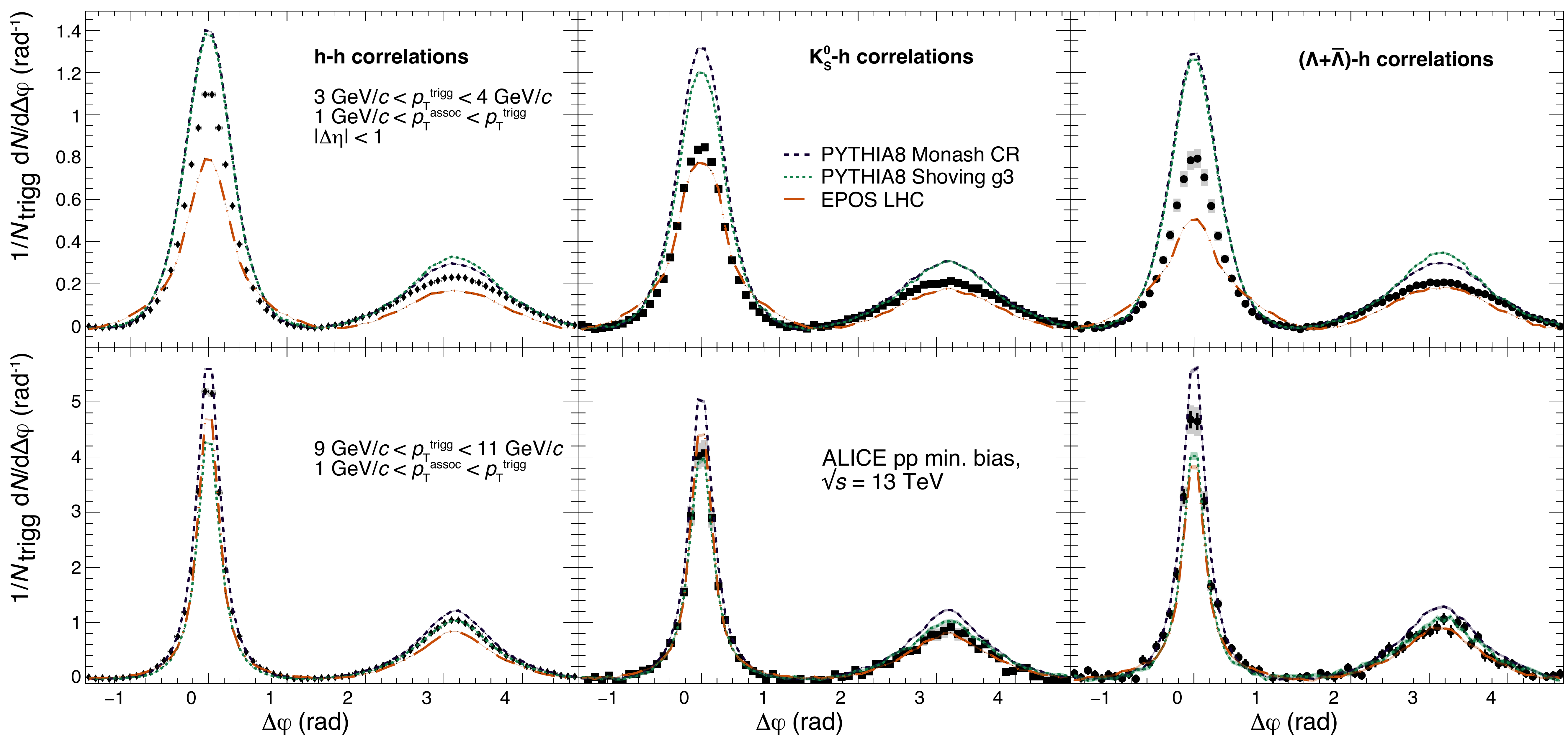}
    \caption{\dphi projection of the h-h (left), \K-h (middle) and \lam-h (right) correlation functions compared with MC event generators for low (top) and high (bottom) \pttrig. Error bars and colored boxes represent statistical and systematic uncertainties, respectively. The data are compared with MC event generators.
    \label{fig:dphi_proj}}
\end{figure}

 The \dphi projections of the correlation functions for the three different trigger particles are shown for two \pttrig intervals,
 $3< \pttrig <4$ \gevc and $9<\ \pttrig <11$ \gevc, in Fig.~\ref{fig:dphi_proj}. 
 Included are also the correlation functions predicted by MC event generators widely used by the LHC collaborations: PYTHIA8 with the~standard Monash tune, which includes colour re-connection as final-state effects~\cite{Skands:2014pea}, PYTHIA8 Monash tune with shoving~\cite{Bierlich:2017vhg} and EPOS LHC~\cite{Pierog:2013ria}. 
 It is important to note that the~PYTHIA 8 Monash and EPOS LHC tunings were based on single-particle spectra and underlying event observables, but did not include particle correlations in azimuth. 
 The shoving strength parameter g is here set to g=3 and in addition the upper \pt cut for the shoving mechanism is turned off.
  None of the models describes quantitatively the~correlation functions consistently for the three trigger particle species. For the low \pttrig, both PYTHIA8 tunes overestimate the peaks on the near- and away-side, while EPOS LHC underestimates them significantly for all trigger particles except for the \K. In the high \pttrig interval, the~shoving tune of PYTHIA8 is underestimating the near-side peak for all trigger particles except for \K-h case, but describes well the away-side peak. The~description of EPOS LHC and PYTHIA8 Monash is similar in both \pttrig intervals, where EPOS LHC is underestimating both peaks of h-h and \lam-h correlation functions and can reasonably well describe the \dphi projection of the \K-h correlation function. PYTHIA8 Monash tune overestimates both peaks for all three types of correlation functions.

The per-trigger yields, obtained by integrating the \dphi projections of the correlation function for the~intervals $|\dphi|<0.9$ (near-side) and $|\dphi-\pi|<1.4$ (away-side) are studied as a function of \pttrig, \ptassoc and event-activity class for the three trigger particle species and compared with the MC event generators. The results are described in the following sections. 

\subsection{Per-trigger yields}

\begin{figure}[hbt!]
    \centering
    \includegraphics[width=0.9\textwidth]{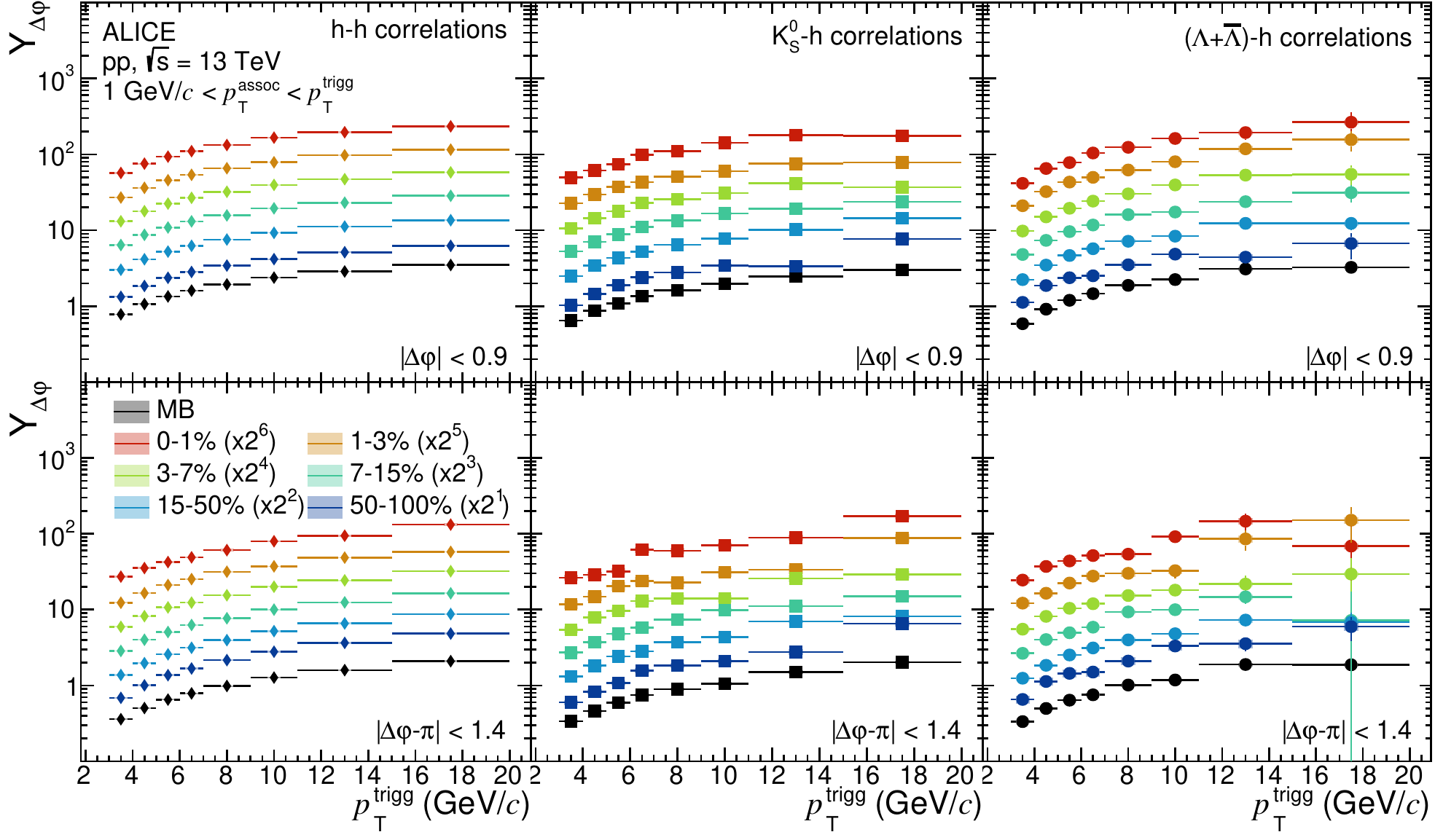}
    \caption{Per-trigger yields of h-h (left), \K-h (middle) and \lam-h (right) correlation functions as a function of \pttrig on the near-side (upper row) and away-side (lower row) for different multiplicity classes. For visibility, the~values in the various event classes are scaled with the factors indicated in the legend. Error bars and colored boxes represent
statistical and systematic uncertainties, respectively, which are in most cases within the data points.}
    \label{fig:yields_pttrigg}
\end{figure}

The per-trigger yields are shown in Fig.~\ref{fig:yields_pttrigg} as a function of the \pttrig for different event-activity classes. An~increasing trend with \pttrig is observed, as expected, as higher energetic jets have increasingly more  associated particles.

\begin{figure}[hbt!]
    \centering
        \includegraphics[width=0.9\textwidth]{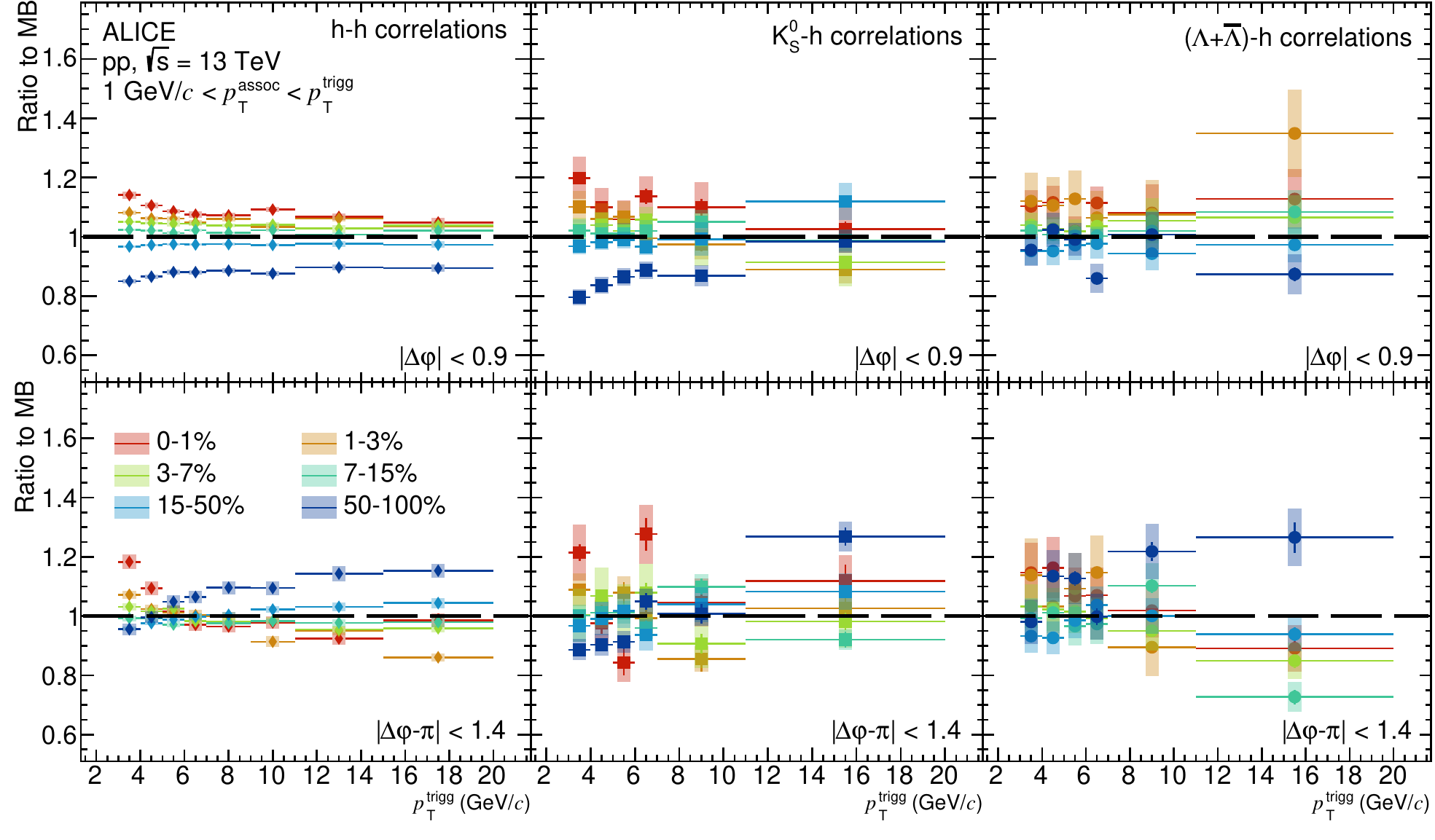}
    \caption{Ratio of the per-trigger yields in different multiplicity classes to the corresponding minimum bias yield for h-h (left), \K-h (middle) and \lam-h (right) correlations on the near-side (upper row) and away-side (lower row). Error bars and colored boxes represent
statistical and systematic uncertainties, respectively.}
    \label{fig:ratio_to_minBias}
\end{figure}

For a more quantitative inspection of the event-activity dependence, the ratio between the yield in each event class and the yield for minimum bias is given in  Fig.~\ref{fig:ratio_to_minBias}. Different trends are visible for the near- and away-side peaks. A clear multiplicity ordering for the near-side can be observed, events with higher multiplicity exhibiting the highest yields.
This behaviour is most obvious for the h-h correlations, but it is significant on the near-side also for the \vzero-triggered yields.

\begin{figure}[hbt!]
    \centering
    \includegraphics[width=0.9\textwidth]{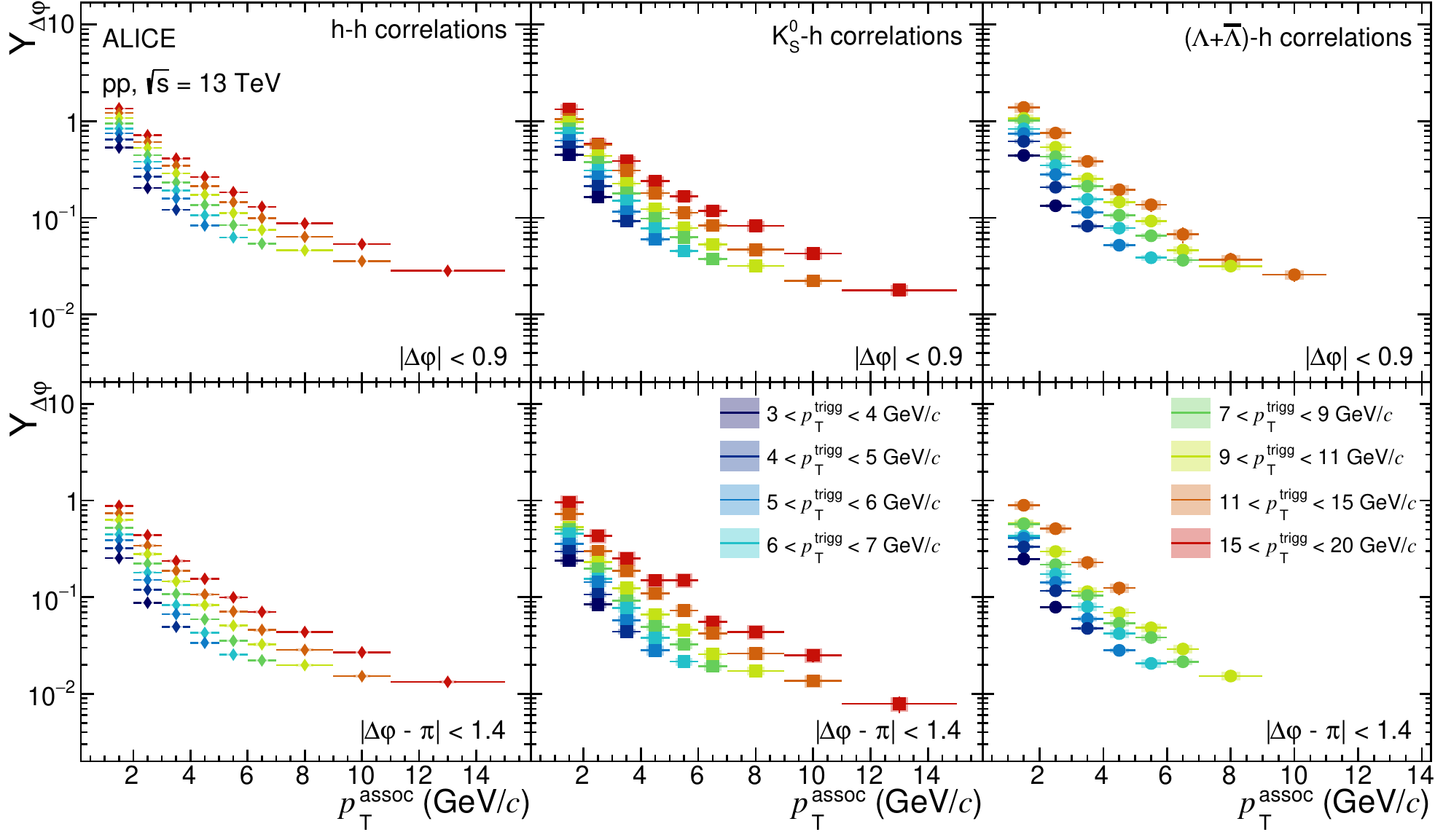}
    \caption{Per-trigger yields of h-h (left), \K-h (middle) and \lam-h (right) correlation functions as a function of \ptassoc on the near-side (upper row) and away-side (lower row) for different \pttrig intervals. Error bars and colored boxes represent statistical and systematic uncertainties, respectively. }
    \label{fig:yields_ptassoc}
\end{figure}

For the away-side this ordering is reverted, in particular with the increase of the \pt of the trigger particle. 
 Given the uncertainties, this trend is less significant for the \K-h and \lam-h correlations. The~finding is qualitatively reproduced in PYTHIA8 simulations and can be understood considering that the location of the away-side jet is not fixed in $\eta$. 
The requirement of a low (high) multiplicity in the V0 detectors at larger rapidity, biases the events towards configurations where the away-side jet is within (outside) the central acceptance, thus increasing (decreasing) the per-trigger yield at the~away-side.   
The~near-side jet is by construction in the central pseudorapidity acceptance, though the multiplicity measurement in the V0 detectors may still be influenced by long range correlations (flow-like) and fragmentation biases. 
This jet bias, although expected and roughly understood, is interesting, as it gives insights on particle production mechanisms. It is further investigated through the comparison with model predictions and through normalisation of the yields for \K-h and \lam-h to those for h-h.

The \ptassoc per-trigger yield spectra for different \pttrig intervals are shown in Fig.~\ref{fig:yields_ptassoc}. A decreasing trend with \ptassoc is observed, as expected, as the probability for creation of associated particle with high \pt decreases with \pt. Note, that due to the kinematic requirement \ptassoc$<$\pttrig, the yields triggered with lower \pt particle do not cover the full \ptassoc range.

\subsection{Comparison with model predictions}

\begin{figure}[hbt!]
    \centering
    \includegraphics[width=0.9\textwidth]{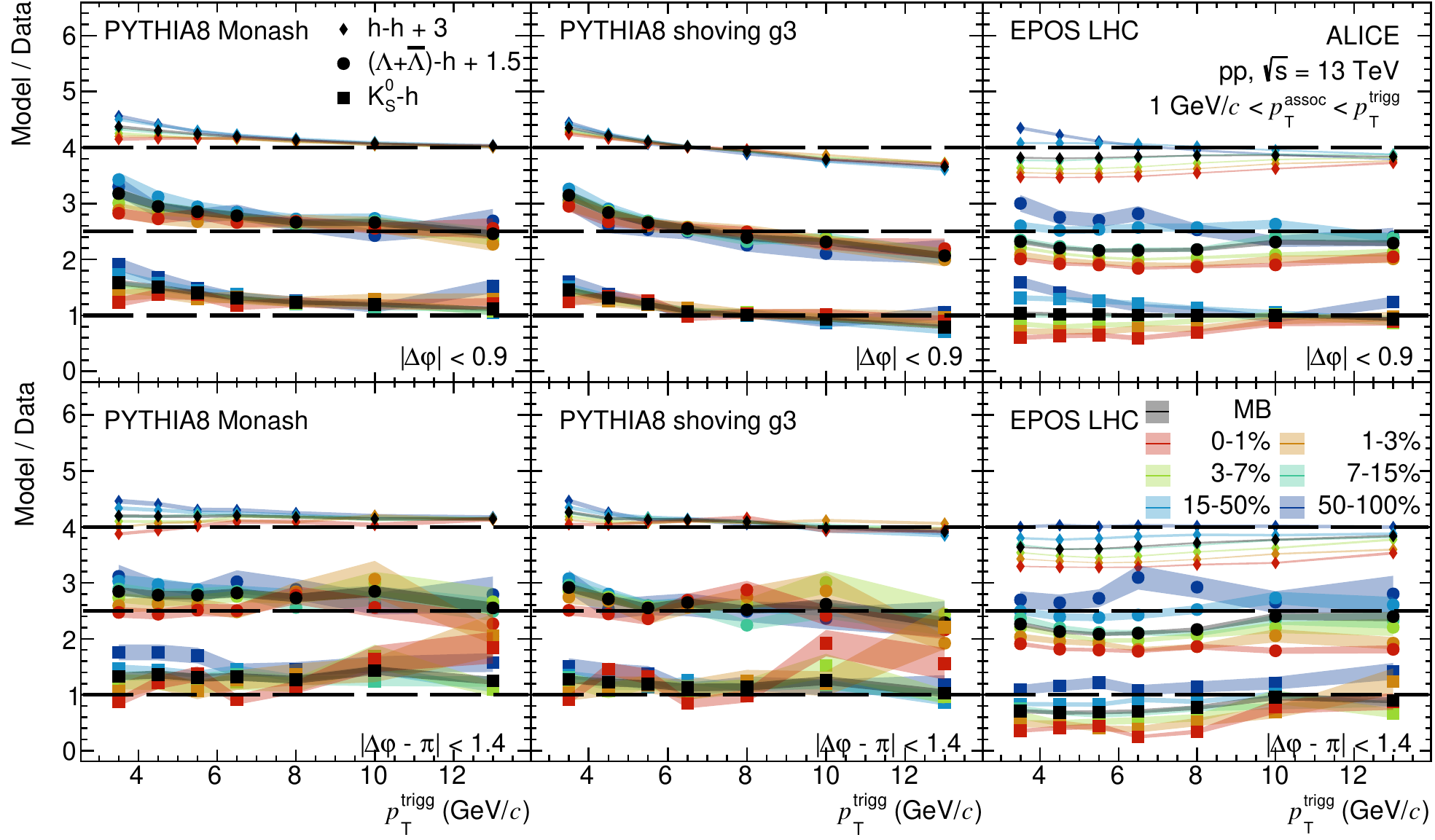}
    \caption{Models to data ratio of integrated per-trigger yields as a function of \pttrig for different multiplicity classes on the near-side (top panels) and away-side (bottom panels) for three MC event generators: PYTHIA8 Monash with colour re-connection (left), PYTHIA8 with shoving (middle) and EPOS LHC (right). Note, for plotting purpose, the~arbitrary shifts for the~vertical axis are given in the legend. The bands represent combined systematic and statistical uncertainties from data and models. }
    \label{fig:MCratio_Yields_pttrigg}
\end{figure}

The ratios of model predictions to data for the integrated per-trigger yields are shown in Fig.~\ref{fig:MCratio_Yields_pttrigg} for the~near- and away-side yields as a function of \pttrig for minimum bias events and event classes selected on their activity. The first finding is that the trends in the model descriptions of the data are largely the same for h-h, \K-h and \lam-h correlations.
The event-activity dependence is reproduced by PYTHIA8 calculations, for both the standard tune and with shoving, the latter clearly offering a better description. The dependence is not reproduced  by EPOS LHC.

For the near-side, PYTHIA8 with colour re-connection describes well the yields for high \pt trigger particles while it overestimates the yields at low-\pt. This is valid for all multiplicity classes and holds for the three species of trigger particle. This is a consequence  of the good description  of the hard QCD processes in PYTHIA8, which do not depend on the event multiplicity. 
The shoving mechanism improves the intermediate sector, at the price of degrading the harder sector. It is apparent that allowing shoving for large \pt, as done for the simulations presented here, is not entirely physical.
The medium-hard and softer regime remains a challenge and is a subject of current theoretical attention~\cite{Bierlich:2020naj}. For the away-side, the~shoving mechanism describes the data better than the standard PYTHIA8.
In EPOS LHC the~event-activity dependence seen in data is not described well: a stronger event-activity dependence is observed in the model, which underpredicts the data for higher event multiplicities and either overpredicts it (for the~near-side) or describes it well (for the~away-side) for the lower event-activity classes.

\begin{figure}[hbt!]
    \centering
    \includegraphics[width=0.9\textwidth]{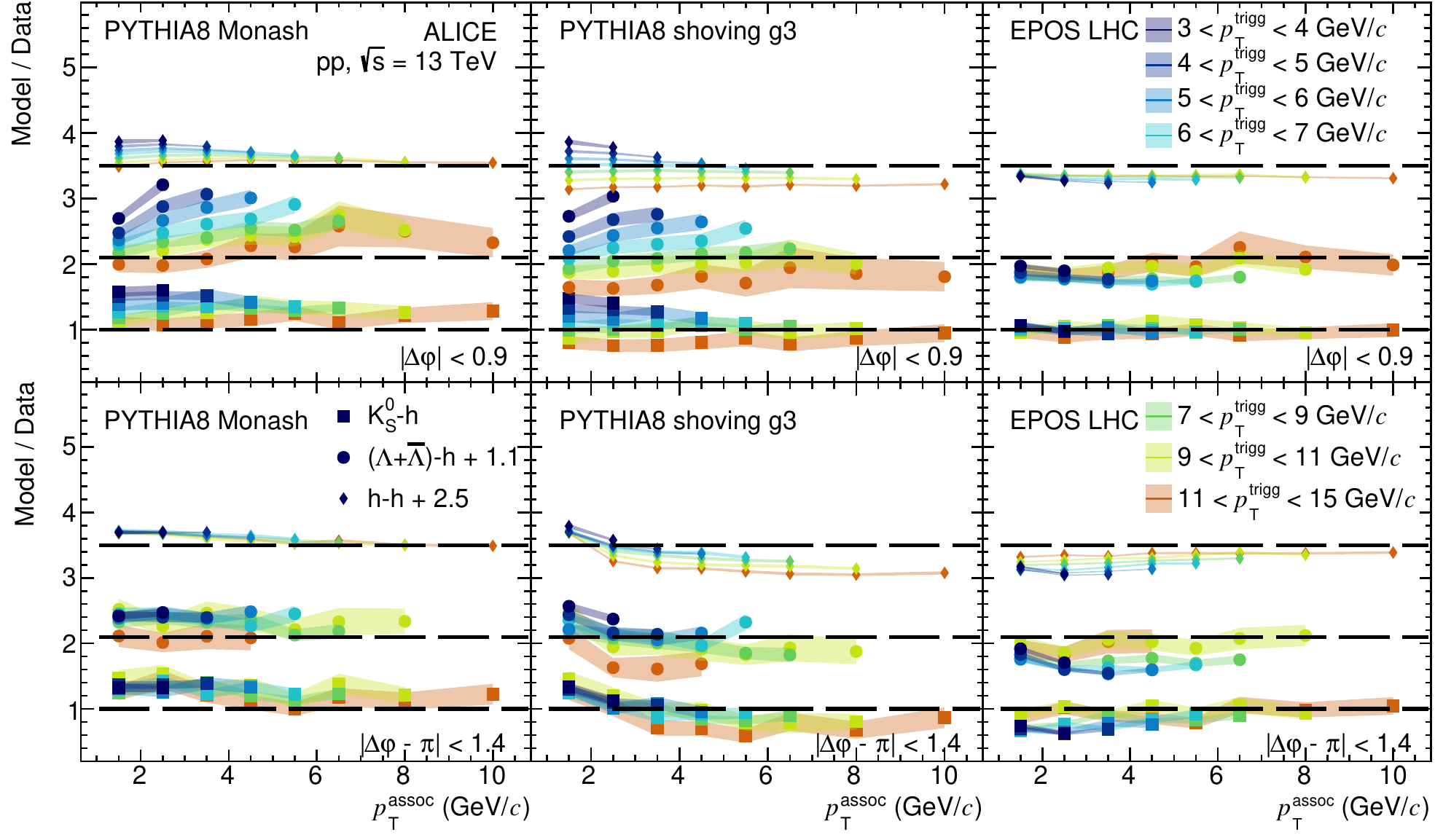}
    \caption{Models to data ratio of integrated per-trigger yields for minimum bias collisions as a function of \ptassoc for different \pttrig intervals on the near-side (top panels) and away-side (bottom panels) for three MC models: PYTHIA8 Monash with colour re-connection (left), PYTHIA8 with shoving (middle) and EPOS LHC (right). Note that for plotting purpose, the~arbitrary shifts for the vertical axis are given in the legend. The bands represent combined systematical and statistical uncertainties for data and statistical for models. }
    \label{fig:MCratio_Yields_ptassoc}
\end{figure}

The features of the data-model comparison are dramatically different in Fig.~\ref{fig:MCratio_Yields_ptassoc}, where the model to data per-trigger yield ratio is shown as a function of \ptassoc for minimum bias collisions. Here, the~trigger-particle dependence is rather different in the models.
PYTHIA8 predicts, for both the standard settings and for the shoving mechanism, in particular for the near-side and in dependence of \pttrig, a much stronger dependence on the trigger particle than seen in the data, while EPOS LHC describes the data well.
For high values of \pttrig, the data are described well in the whole \ptassoc range. This reflects that the~full fragmentation process in a hard scattering is well modelled. 
It is important to note that PYTHIA8 Monash describes quite well the measured \pt dependence of the ratio $\lam/\K$ in jets with $\pt~>~10$~\gevc. However, the ratio in MB collisions is not described well, in particular for $\pt<4$~\gevc~\cite{Acharya:2021oaa}.

It is interesting to examine the differences between PYTHIA8 with the~standard settings and with the~shoving mechanism. 
Clearly, the shoving mechanism describes the medium-hard processes better, corresponding roughly to $3<\pttrig<5$ \gevc and $1<\ptassoc<5$ \gevc, while the harder processes are underestimated. This is visible in particular in Fig.~\ref{fig:MCratio_Yields_pttrigg} for all trigger-particle species for different event-activity classes, while in case of the \ptassoc dependence shown in Fig.~\ref{fig:MCratio_Yields_ptassoc} the shoving mechanism leads to a~poorer description of the data compared to the standard settings. 

The EPOS LHC model gives a rather different description of the data compared to PYTHIA8. In contrast to the rather poor description of the \pttrig  and event-class dependence (Fig.~\ref{fig:MCratio_Yields_pttrigg}) the dependence on \ptassoc  (Fig.~\ref{fig:MCratio_Yields_ptassoc}) is rather well described for different \pttrig intervals and for the three trigger-particle species.

\subsection{Comparison of hadron- and strangeness-triggered yields}

\begin{figure}[hbt]
    \centering
    \begin{subfigure}
    \centering
    \includegraphics[width=0.49\textwidth]{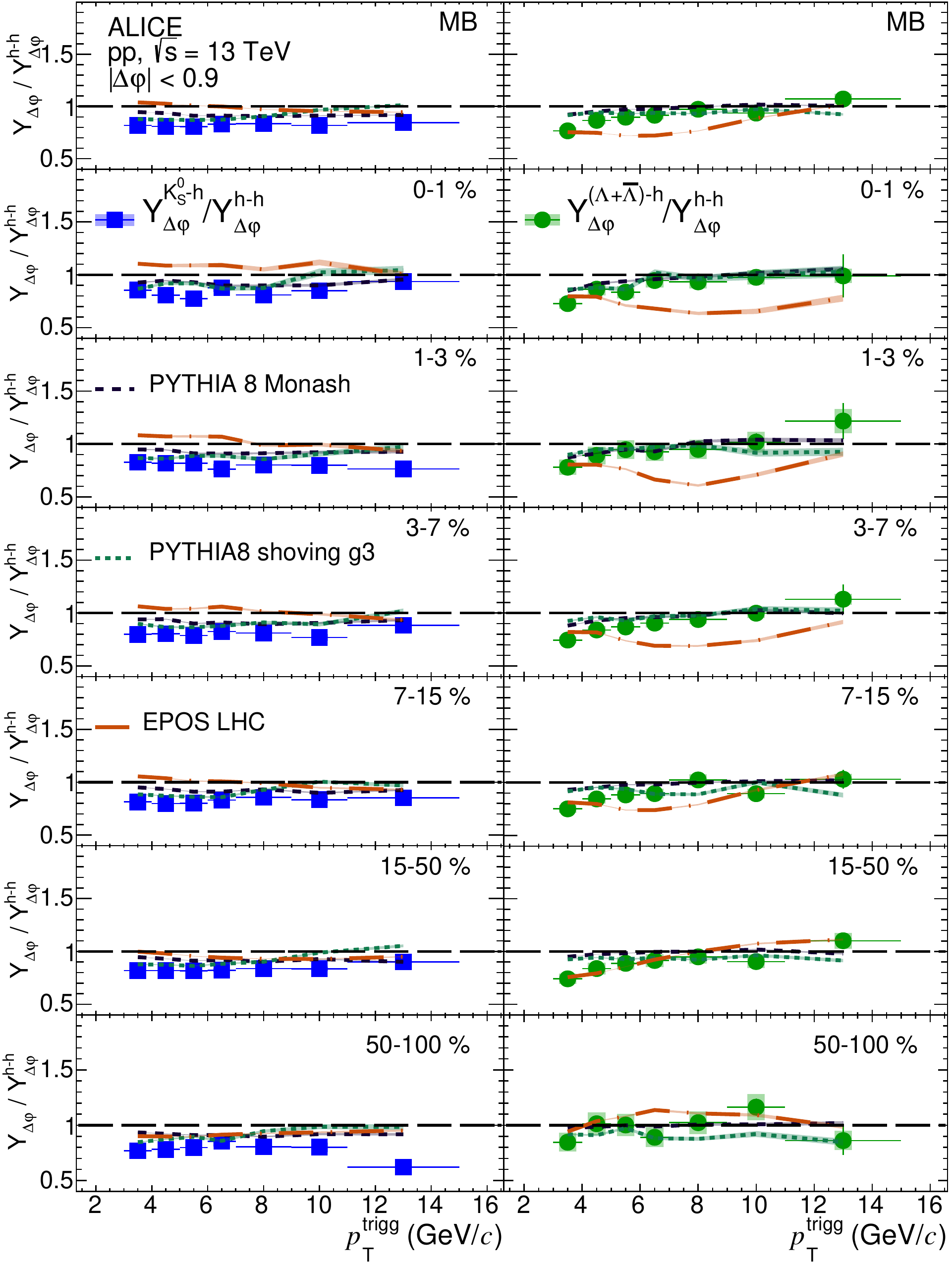}
    \includegraphics[width=0.49\textwidth]{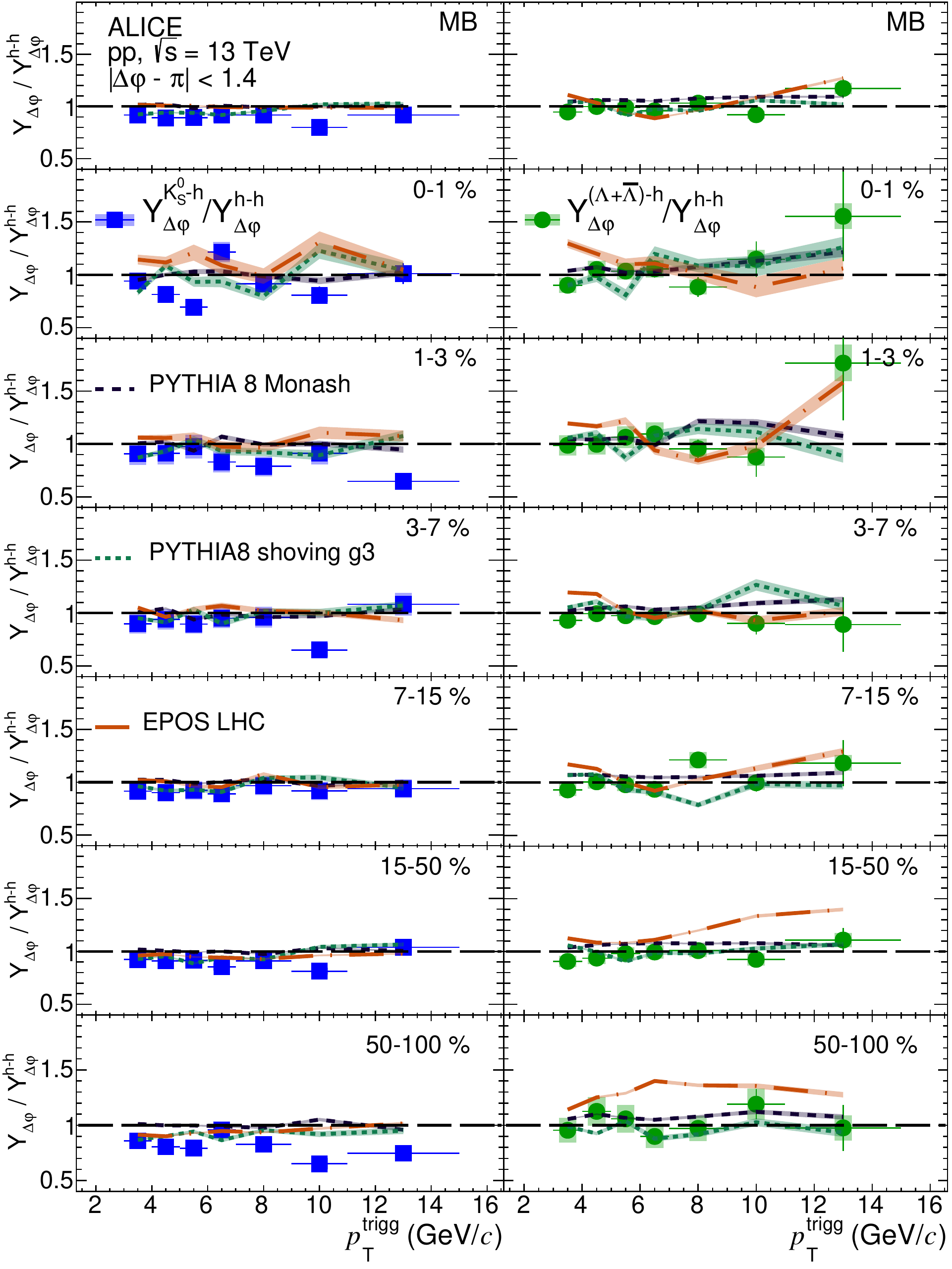}
    \end{subfigure}
    \caption{ Ratios of integrated per-trigger yield of \K-h (left column) or \lam-h (right column) to h-h as a function of \pttrig, for the near-side in the left plot and for the away-side in the right plot, for different event multiplicity classes. Error bars and colored boxes represent statistical and systematic uncertainties, respectively. The bands around model curves stand for their statistical uncertainty.}
    \label{fig:Ratio_to_hh_pttrigg}
\end{figure}

\begin{figure}[hbt]
    \centering
    \begin{subfigure}
    \centering
    \includegraphics[width=0.49\textwidth]{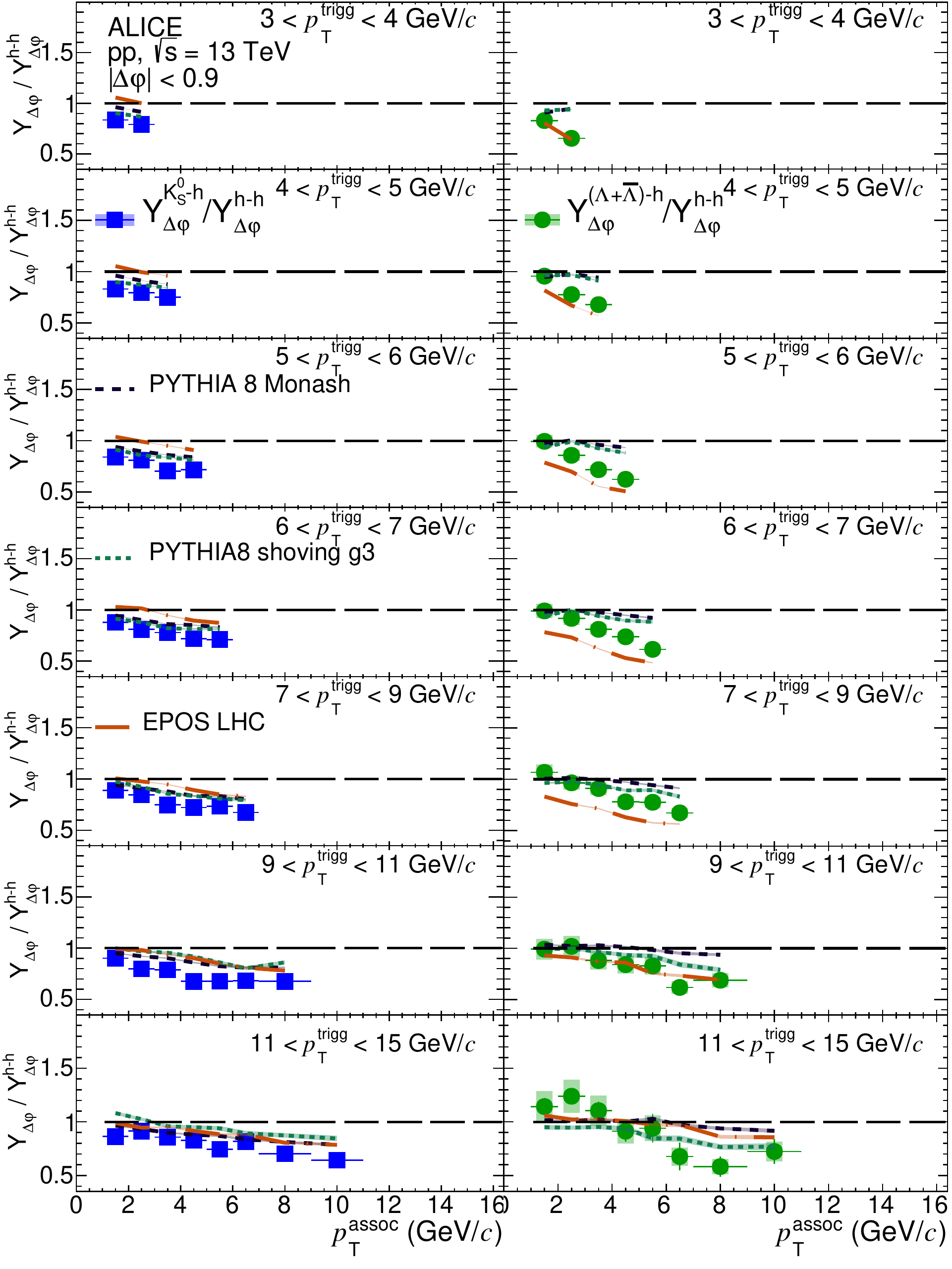}
    \end{subfigure}
    \begin{subfigure}
    \centering
    \includegraphics[width=0.49\textwidth]{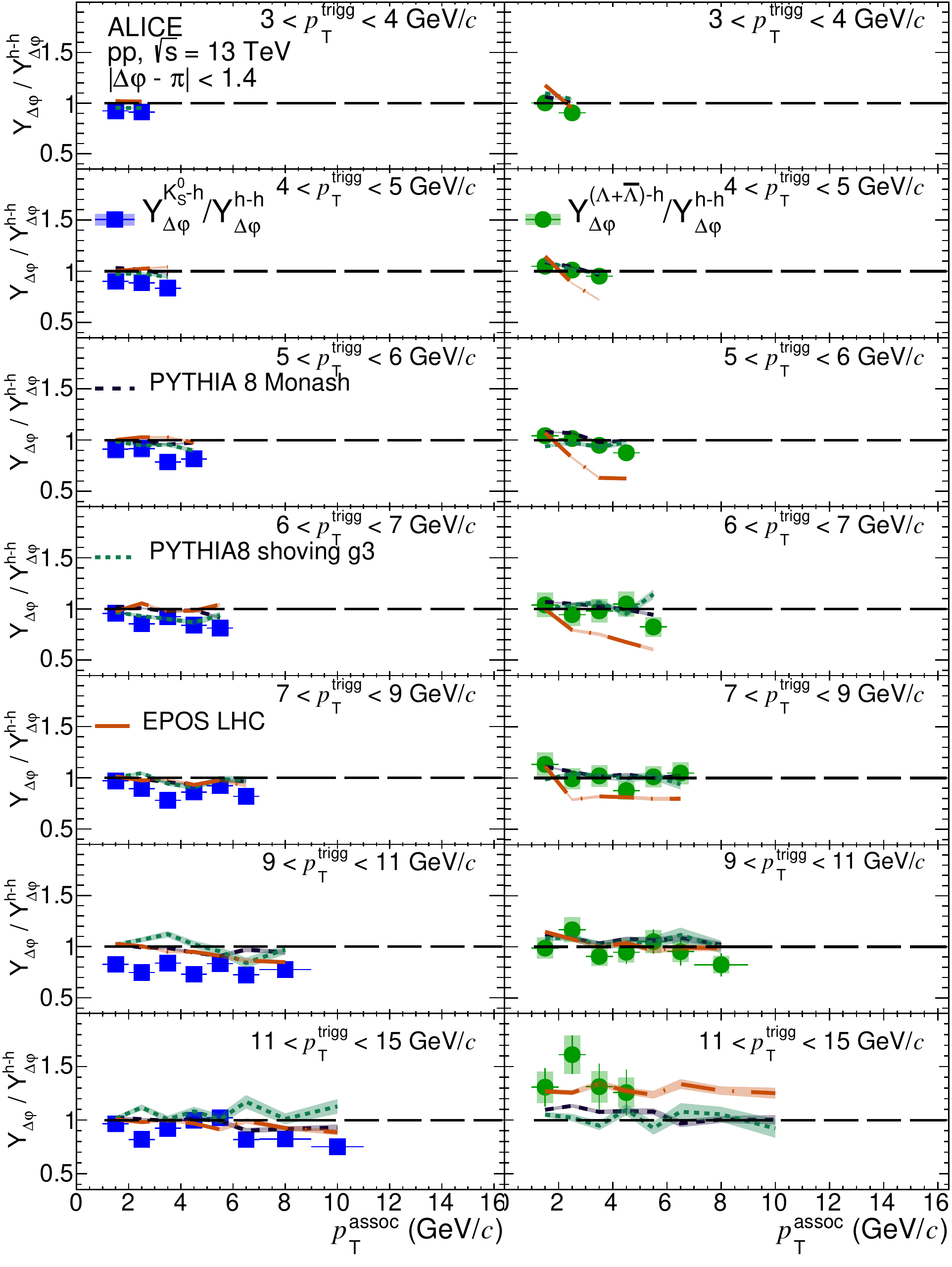}
    \end{subfigure}
    \caption{Per-trigger yield ratios of \K-h (left column) or \lam-h (right column) to h-h as a function of \ptassoc, for the near-side in the left plot and for the away-side in the right plot, for different \pttrig intervals in the MB sample. Error bars and colored boxes represent statistical and systematic uncertainties, respectively. The bands around model curves stand for their statistical uncertainty.}
    \label{fig:Ratio_to_hh_ptassoc}
\end{figure}

In order to compare more quantitatively the correlation yields for different trigger-particle species, ratios to the per-trigger yields for h-h are calculated. These ratios are shown as a function of \pttrig in Fig.~\ref{fig:Ratio_to_hh_pttrigg} for the near- and away-side, while Fig.~\ref{fig:Ratio_to_hh_ptassoc}
shows them as a function of \ptassoc. The data indicate that the~differences between charged-hadron (a sample dominated by pions) triggered yields and either \K or \lam triggered yields are small and have a weak dependence on the event-activity class. 
All three models describe the trends seen in the ratios better than it was the case for yields itself, a consequence of the earlier finding that
the model-data disagreements are very similar for the three trigger-particle species.

At closer examination, one observes that a difference between different trigger-particle species is visible for the near-side. The yield ratios $\rm Y_{\Delta\varphi}^{\K-h}/Y_{\Delta\varphi}^{h-h}$ are smaller than unity and flat with \pttrig in all multiplicity classes, which indicates that jets triggered with \K mesons contain less particles than the~unbiased (inclusive) jets and this feature does not depend on the hardness of the process (\pttrig) or event multiplicity. In contrast to that, the  $\rm Y_{\Delta\varphi}^{(\Lambda+\overline{\Lambda})-h}/Y_{\Delta\varphi}^{h-h}$ ratio increases with \pttrig. Potentially, this could be explained with a bias towards gluon jets, which contain more particles~\cite{Abbiendi:2002} and have relative enhanced production of $\Lambda$ hyperons~\cite{Ackerstaff:1999}. In the right column in the left plot of Fig.~\ref{fig:Ratio_to_hh_ptassoc}, it is visible that this effect in the data is pronounced for the soft part of harder jets (low \ptassoc for high \pttrig). A decreasing trend of the ratio can also be observed with increasing \ptassoc. This suggests that a jet triggered with either a \K or a $\Lambda$ or $\overline{\Lambda}$ particle has a smaller amount of associated particles with higher \pt than a~jet triggered with an~unidentified hadron with the~same \pt. This decreasing trend is reproduced by the considered models. 

\begin{figure}[hbt!]
    \centering
    \begin{subfigure}
    \centering
    \includegraphics[width=0.49\textwidth]{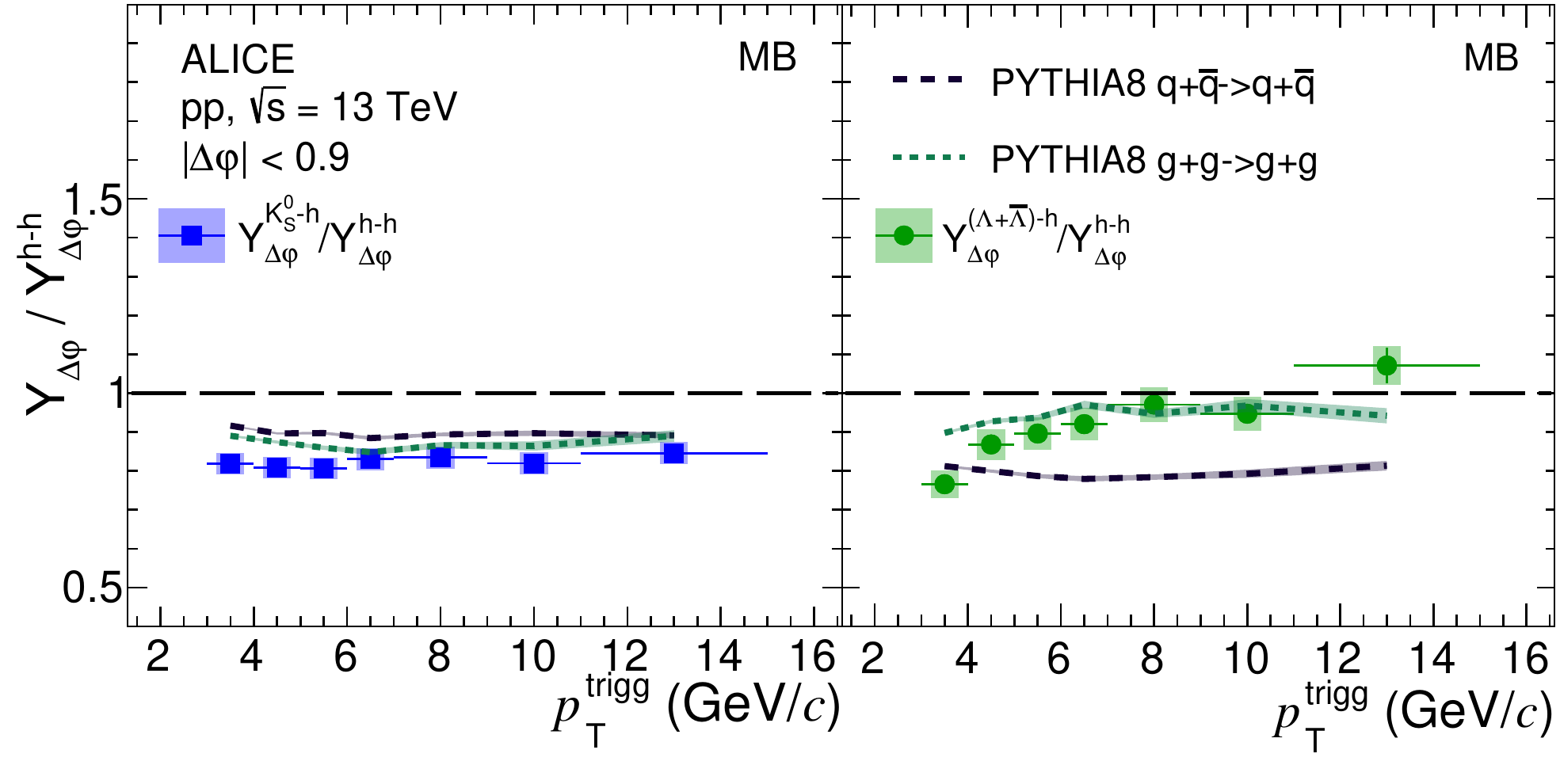}
    \end{subfigure}
    \begin{subfigure}
    \centering
    \includegraphics[width=0.49\textwidth]{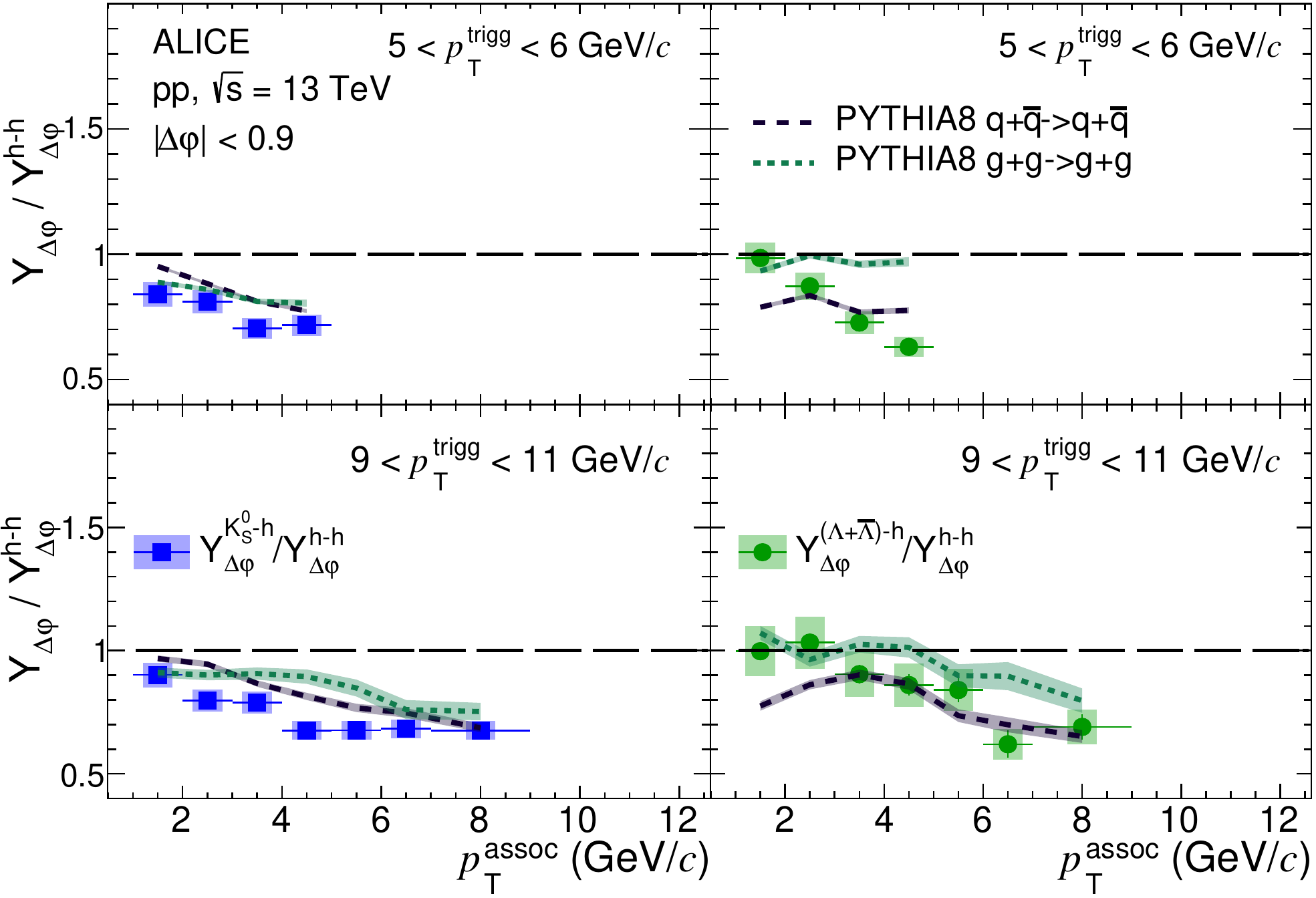}
    \end{subfigure}
    \caption{Per-trigger yield ratios of \K-h (left column) or \lam-h (right column) to h-h at the near-side as a function of \pttrig in the left plot and as a function of \ptassoc in the right plot, compared with the PYTHIA8 calculation of the quark and gluon jets. Error bars and colored boxes represent statistical and systematic uncertainties, respectively. The bands around model curves stand for their statistical uncertainty.}
    \label{fig:Ratio_to_hh_quark_gluon}
\end{figure}

To check if the difference in the ratios for the near side triggered with \K and \lam can be caused by the~differences between quark and gluon jets, a separated PYTHIA 8 study was performed, selecting exclusive hard processes containing only quarks ($q+\overline{q}\rightarrow q+\overline{q}$) or gluons ($g+g\rightarrow g+g$) in the final state. The results of this study are shown in Fig.~\ref{fig:Ratio_to_hh_quark_gluon} for MB collisions 
in comparison with the data. It is visible that the ratio is almost identical for the \K-triggered yields for both quark and gluon jets. In this case, for both processes the ratio is flat as a function of \pttrig and decreasing with \ptassoc. Nevertheless, a~clear difference between the two processes is observed in the ratios of yields triggered with \lam. The~ratio for the gluon-jet process is increasing with \pttrig and is higher, while it is flat and smaller for the~quark jets. A systematic difference is visible also in the \ptassoc dependence, where the ratio from gluon jets is significantly higher than the one from quark jets. Although the~\pt distributions of hadrons in the~simulated exclusive processes may be different than in MB collisions, it is expected that the observed difference is generic and explains at least some of the trends seen in the data.

In contrast, for the away-side, the ratios as a function of \pttrig are compatible with unity within the~uncertainties as shown in the right part of Fig.~\ref{fig:Ratio_to_hh_pttrigg}, thus no dependence on the trigger particle species (\K or \lam) is seen. The deviation of some bins from unity is caused by the fluctuations affecting the underlying event estimation. A similar trend is also observed in the \ptassoc dependence in the right plot of Fig.~\ref{fig:Ratio_to_hh_ptassoc}. This fits to the expectation that there is no bias in the away-side jet, since the associated production of strangeness is not dominated by hard processes, i.e. initial production of a $\mathrm{s\bar{s}}$ pair leading to two strangeness induced jets.  
This is confirmed by the comparison with models, which predict ratios for the~away-side around unity (Fig.~\ref{fig:Ratio_to_hh_pttrigg}).

\section{Summary and conclusions}
\label{sec:conc}

Charged-hadron, \K and \lam triggered correlation functions are measured at the LHC in pp collisions at $\sqrt{s}$ = 13 TeV with the ALICE apparatus.
The integrated per-trigger yields are extracted for both near- and away-side and studied as a function of event multiplicity, \pttrig and \ptassoc. 
A dependence on the~event-multiplicity is observed, which may be explained at the near-side by an interplay of fragmentation and of collective-like effects such as the long-range ridge. For the away-side the primary effects are likely due to a selection bias, where the~away-side jet is within the V0 detector acceptance causing smaller away-side yields for high-multiplicity classes.
This bias was overcome by studying the ratios of yields for \K or \lam as trigger particle over that for inclusive hadrons. For the near-side, different trends were observed for yield ratios triggered by \K and \lam. While the $\rm Y_{\Delta\varphi}^{K_S^0-h}/Y_{\Delta\varphi}^{h-h}$ is flat, the~$\rm Y_{\Delta\varphi}^{(\Lambda+\overline{\Lambda})-h}/Y_{\Delta\varphi}^{h-h}$ increases with \pttrig.
This difference is most pronounced for high \pttrig and low \ptassoc. A PYTHIA8 study of the respective exclusive processes reveals that the difference is caused by the bias towards gluon jets through triggering with $\Lambda$ or $\overline{\Lambda}$ hyperon.
In the \ptassoc dependence of these yield ratios, a~decreasing trend of the ratio can be observed  with increasing \ptassoc, suggesting a smaller amount of associated particles with higher \pt in the jets triggered with a \vzero particle than in jets triggered with an~unidentified hadron with the same \pt. This trend can be reproduced by the considered models.
However, on the~away-side, no clear difference between the trigger-particle species is observed.

All the measured quantities are compared with model predictions.  PYTHIA8 with the standard Monash tune can describe the data for high \pttrig (hard processes) while the shoving mechanism works better at intermediate \pttrig. The comparison of data with the EPOS LHC model reveals that the model  predicts a significantly stronger multiplicity dependence than seen in the data. The  ratios of yields for \K or \lam trigger particles to those for inclusive (charged) hadrons are qualitatively well described by all the considered models, except for EPOS LHC at the near-side for the \lam triggered yield ratios.

\newenvironment{acknowledgement}{\relax}{\relax} 
\begin{acknowledgement}   
\section*{Acknowledgements}
The ALICE Collaboration would like to thank Christian Bierlich for providing the PYTHIA8 String Shoving configurations and discussions around this and Tanguy Pierog and Klaus Werner for helpful discussions around the EPOS LHC model.  

The ALICE Collaboration would like to thank all its engineers and technicians for their invaluable contributions to the construction of the experiment and the CERN accelerator teams for the outstanding performance of the LHC complex.
The ALICE Collaboration gratefully acknowledges the resources and support provided by all Grid centres and the Worldwide LHC Computing Grid (WLCG) collaboration.
The ALICE Collaboration acknowledges the following funding agencies for their support in building and running the ALICE detector:
A. I. Alikhanyan National Science Laboratory (Yerevan Physics Institute) Foundation (ANSL), State Committee of Science and World Federation of Scientists (WFS), Armenia;
Austrian Academy of Sciences, Austrian Science Fund (FWF): [M 2467-N36] and Nationalstiftung f\"{u}r Forschung, Technologie und Entwicklung, Austria;
Ministry of Communications and High Technologies, National Nuclear Research Center, Azerbaijan;
Conselho Nacional de Desenvolvimento Cient\'{\i}fico e Tecnol\'{o}gico (CNPq), Financiadora de Estudos e Projetos (Finep), Funda\c{c}\~{a}o de Amparo \`{a} Pesquisa do Estado de S\~{a}o Paulo (FAPESP) and Universidade Federal do Rio Grande do Sul (UFRGS), Brazil;
Ministry of Education of China (MOEC) , Ministry of Science \& Technology of China (MSTC) and National Natural Science Foundation of China (NSFC), China;
Ministry of Science and Education and Croatian Science Foundation, Croatia;
Centro de Aplicaciones Tecnol\'{o}gicas y Desarrollo Nuclear (CEADEN), Cubaenerg\'{\i}a, Cuba;
Ministry of Education, Youth and Sports of the Czech Republic, Czech Republic;
The Danish Council for Independent Research | Natural Sciences, the VILLUM FONDEN and Danish National Research Foundation (DNRF), Denmark;
Helsinki Institute of Physics (HIP), Finland;
Commissariat \`{a} l'Energie Atomique (CEA) and Institut National de Physique Nucl\'{e}aire et de Physique des Particules (IN2P3) and Centre National de la Recherche Scientifique (CNRS), France;
Bundesministerium f\"{u}r Bildung und Forschung (BMBF) and GSI Helmholtzzentrum f\"{u}r Schwerionenforschung GmbH, Germany;
General Secretariat for Research and Technology, Ministry of Education, Research and Religions, Greece;
National Research, Development and Innovation Office, Hungary;
Department of Atomic Energy Government of India (DAE), Department of Science and Technology, Government of India (DST), University Grants Commission, Government of India (UGC) and Council of Scientific and Industrial Research (CSIR), India;
Indonesian Institute of Science, Indonesia;
Istituto Nazionale di Fisica Nucleare (INFN), Italy;
Institute for Innovative Science and Technology , Nagasaki Institute of Applied Science (IIST), Japanese Ministry of Education, Culture, Sports, Science and Technology (MEXT) and Japan Society for the Promotion of Science (JSPS) KAKENHI, Japan;
Consejo Nacional de Ciencia (CONACYT) y Tecnolog\'{i}a, through Fondo de Cooperaci\'{o}n Internacional en Ciencia y Tecnolog\'{i}a (FONCICYT) and Direcci\'{o}n General de Asuntos del Personal Academico (DGAPA), Mexico;
Nederlandse Organisatie voor Wetenschappelijk Onderzoek (NWO), Netherlands;
The Research Council of Norway, Norway;
Commission on Science and Technology for Sustainable Development in the South (COMSATS), Pakistan;
Pontificia Universidad Cat\'{o}lica del Per\'{u}, Peru;
Ministry of Education and Science, National Science Centre and WUT ID-UB, Poland;
Korea Institute of Science and Technology Information and National Research Foundation of Korea (NRF), Republic of Korea;
Ministry of Education and Scientific Research, Institute of Atomic Physics and Ministry of Research and Innovation and Institute of Atomic Physics, Romania;
Joint Institute for Nuclear Research (JINR), Ministry of Education and Science of the Russian Federation, National Research Centre Kurchatov Institute, Russian Science Foundation and Russian Foundation for Basic Research, Russia;
Ministry of Education, Science, Research and Sport of the Slovak Republic, Slovakia;
National Research Foundation of South Africa, South Africa;
Swedish Research Council (VR) and Knut \& Alice Wallenberg Foundation (KAW), Sweden;
European Organization for Nuclear Research, Switzerland;
Suranaree University of Technology (SUT), National Science and Technology Development Agency (NSDTA) and Office of the Higher Education Commission under NRU project of Thailand, Thailand;
Turkish Energy, Nuclear and Mineral Research Agency (TENMAK), Turkey;
National Academy of  Sciences of Ukraine, Ukraine;
Science and Technology Facilities Council (STFC), United Kingdom;
National Science Foundation of the United States of America (NSF) and United States Department of Energy, Office of Nuclear Physics (DOE NP), United States of America.    
\end{acknowledgement}


\bibliographystyle{utphys}   
\bibliography{correl}

\newpage 
\appendix
\section{The ALICE Collaboration} 
\label{app:collab}
\small
\begin{flushleft}

S.~Acharya$^{\rm 141}$, 
D.~Adamov\'{a}$^{\rm 96}$, 
A.~Adler$^{\rm 74}$, 
G.~Aglieri Rinella$^{\rm 34}$, 
M.~Agnello$^{\rm 30}$, 
N.~Agrawal$^{\rm 54}$, 
Z.~Ahammed$^{\rm 141}$, 
S.~Ahmad$^{\rm 16}$, 
S.U.~Ahn$^{\rm 76}$, 
I.~Ahuja$^{\rm 38}$
% ...

\section*{Affiliation notes}

$^{\rm I}$ Deceased\\
$^{\rm II}$ Also at: Italian National Agency for New Technologies, Energy and Sustainable Economic Development (ENEA), Bologna, Italy\\
$^{\rm III}$ Also at: Dipartimento DET del Politecnico di Torino, Turin, Italy\\
$^{\rm IV}$ Also at: M.V. Lomonosov Moscow State University, D.V. Skobeltsyn Institute of Nuclear, Physics, Moscow, Russia\\
$^{\rm V}$ Also at: Department of Applied Physics, Aligarh Muslim University, Aligarh, India
\\
$^{\rm VI}$ Also at: Institute of Theoretical Physics, University of Wroclaw, Poland\\

\section*{Collaboration Institutes}

$^{1}$ A.I. Alikhanyan National Science Laboratory (Yerevan Physics Institute) Foundation, Yerevan, Armenia\\
$^{2}$ AGH University of Science and Technology, Cracow, Poland\\
$^{3}$ Bogolyubov Institute for Theoretical Physics, National Academy of Sciences of Ukraine, Kiev, Ukraine\\
$^{4}$ Bose Institute, Department of Physics  and Centre for Astroparticle Physics and Space Science (CAPSS), Kolkata, India\\
$^{5}$ Budker Institute for Nuclear Physics, Novosibirsk, Russia\\
$^{6}$ California Polytechnic State University, San Luis Obispo, California, United States\\
$^{7}$ Central China Normal University, Wuhan, China\\
$^{8}$ Centro de Aplicaciones Tecnol\'{o}gicas y Desarrollo Nuclear (CEADEN), Havana, Cuba\\
$^{9}$ Centro de Investigaci\'{o}n y de Estudios Avanzados (CINVESTAV), Mexico City and M\'{e}rida, Mexico\\
$^{10}$ Chicago State University, Chicago, Illinois, United States\\
$^{11}$ China Institute of Atomic Energy, Beijing, China\\
$^{12}$ Chungbuk National University, Cheongju, Republic of Korea\\
$^{13}$ Comenius University Bratislava, Faculty of Mathematics, Physics and Informatics, Bratislava, Slovakia\\
$^{14}$ COMSATS University Islamabad, Islamabad, Pakistan\\
$^{15}$ Creighton University, Omaha, Nebraska, United States\\
$^{16}$ Department of Physics, Aligarh Muslim University, Aligarh, India\\
$^{17}$ Department of Physics, Pusan National University, Pusan, Republic of Korea\\
$^{18}$ Department of Physics, Sejong University, Seoul, Republic of Korea\\
$^{19}$ Department of Physics, University of California, Berkeley, California, United States\\
$^{20}$ Department of Physics, University of Oslo, Oslo, Norway\\
$^{21}$ Department of Physics and Technology, University of Bergen, Bergen, Norway\\
$^{22}$ Dipartimento di Fisica dell'Universit\`{a} and Sezione INFN, Cagliari, Italy\\
$^{23}$ Dipartimento di Fisica dell'Universit\`{a} and Sezione INFN, Trieste, Italy\\
$^{24}$ Dipartimento di Fisica dell'Universit\`{a} and Sezione INFN, Turin, Italy\\
$^{25}$ Dipartimento di Fisica e Astronomia dell'Universit\`{a} and Sezione INFN, Bologna, Italy\\
$^{26}$ Dipartimento di Fisica e Astronomia dell'Universit\`{a} and Sezione INFN, Catania, Italy\\
$^{27}$ Dipartimento di Fisica e Astronomia dell'Universit\`{a} and Sezione INFN, Padova, Italy\\
$^{28}$ Dipartimento di Fisica e Nucleare e Teorica, Universit\`{a} di Pavia, Pavia, Italy\\
$^{29}$ Dipartimento di Fisica `E.R.~Caianiello' dell'Universit\`{a} and Gruppo Collegato INFN, Salerno, Italy\\
$^{30}$ Dipartimento DISAT del Politecnico and Sezione INFN, Turin, Italy\\
$^{31}$ Dipartimento di Scienze e Innovazione Tecnologica dell'Universit\`{a} del Piemonte Orientale and INFN Sezione di Torino, Alessandria, Italy\\
$^{32}$ Dipartimento di Scienze MIFT, Universit\`{a} di Messina, Messina, Italy\\
$^{33}$ Dipartimento Interateneo di Fisica `M.~Merlin' and Sezione INFN, Bari, Italy\\
$^{34}$ European Organization for Nuclear Research (CERN), Geneva, Switzerland\\
$^{35}$ Faculty of Electrical Engineering, Mechanical Engineering and Naval Architecture, University of Split, Split, Croatia\\
$^{36}$ Faculty of Engineering and Science, Western Norway University of Applied Sciences, Bergen, Norway\\
$^{37}$ Faculty of Nuclear Sciences and Physical Engineering, Czech Technical University in Prague, Prague, Czech Republic\\
$^{38}$ Faculty of Science, P.J.~\v{S}af\'{a}rik University, Ko\v{s}ice, Slovakia\\
$^{39}$ Frankfurt Institute for Advanced Studies, Johann Wolfgang Goethe-Universit\"{a}t Frankfurt, Frankfurt, Germany\\
$^{40}$ Fudan University, Shanghai, China\\
$^{41}$ Gangneung-Wonju National University, Gangneung, Republic of Korea\\
$^{42}$ Gauhati University, Department of Physics, Guwahati, India\\
$^{43}$ Helmholtz-Institut f\"{u}r Strahlen- und Kernphysik, Rheinische Friedrich-Wilhelms-Universit\"{a}t Bonn, Bonn, Germany\\
$^{44}$ Helsinki Institute of Physics (HIP), Helsinki, Finland\\
$^{45}$ High Energy Physics Group,  Universidad Aut\'{o}noma de Puebla, Puebla, Mexico\\
$^{46}$ Hiroshima University, Hiroshima, Japan\\
$^{47}$ Hochschule Worms, Zentrum  f\"{u}r Technologietransfer und Telekommunikation (ZTT), Worms, Germany\\
$^{48}$ Horia Hulubei National Institute of Physics and Nuclear Engineering, Bucharest, Romania\\
$^{49}$ Indian Institute of Technology Bombay (IIT), Mumbai, India\\
$^{50}$ Indian Institute of Technology Indore, Indore, India\\
$^{51}$ Indonesian Institute of Sciences, Jakarta, Indonesia\\
$^{52}$ INFN, Laboratori Nazionali di Frascati, Frascati, Italy\\
$^{53}$ INFN, Sezione di Bari, Bari, Italy\\
$^{54}$ INFN, Sezione di Bologna, Bologna, Italy\\
$^{55}$ INFN, Sezione di Cagliari, Cagliari, Italy\\
$^{56}$ INFN, Sezione di Catania, Catania, Italy\\
$^{57}$ INFN, Sezione di Padova, Padova, Italy\\
$^{58}$ INFN, Sezione di Pavia, Pavia, Italy\\
$^{59}$ INFN, Sezione di Torino, Turin, Italy\\
$^{60}$ INFN, Sezione di Trieste, Trieste, Italy\\
$^{61}$ Inha University, Incheon, Republic of Korea\\
$^{62}$ Institute for Gravitational and Subatomic Physics (GRASP), Utrecht University/Nikhef, Utrecht, Netherlands\\
$^{63}$ Institute for Nuclear Research, Academy of Sciences, Moscow, Russia\\
$^{64}$ Institute of Experimental Physics, Slovak Academy of Sciences, Ko\v{s}ice, Slovakia\\
$^{65}$ Institute of Physics, Homi Bhabha National Institute, Bhubaneswar, India\\
$^{66}$ Institute of Physics of the Czech Academy of Sciences, Prague, Czech Republic\\
$^{67}$ Institute of Space Science (ISS), Bucharest, Romania\\
$^{68}$ Institut f\"{u}r Kernphysik, Johann Wolfgang Goethe-Universit\"{a}t Frankfurt, Frankfurt, Germany\\
$^{69}$ Instituto de Ciencias Nucleares, Universidad Nacional Aut\'{o}noma de M\'{e}xico, Mexico City, Mexico\\
$^{70}$ Instituto de F\'{i}sica, Universidade Federal do Rio Grande do Sul (UFRGS), Porto Alegre, Brazil\\
$^{71}$ Instituto de F\'{\i}sica, Universidad Nacional Aut\'{o}noma de M\'{e}xico, Mexico City, Mexico\\
$^{72}$ iThemba LABS, National Research Foundation, Somerset West, South Africa\\
$^{73}$ Jeonbuk National University, Jeonju, Republic of Korea\\
$^{74}$ Johann-Wolfgang-Goethe Universit\"{a}t Frankfurt Institut f\"{u}r Informatik, Fachbereich Informatik und Mathematik, Frankfurt, Germany\\
$^{75}$ Joint Institute for Nuclear Research (JINR), Dubna, Russia\\
$^{76}$ Korea Institute of Science and Technology Information, Daejeon, Republic of Korea\\
$^{77}$ KTO Karatay University, Konya, Turkey\\
$^{78}$ Laboratoire de Physique des 2 Infinis, Ir\`{e}ne Joliot-Curie, Orsay, France\\
$^{79}$ Laboratoire de Physique Subatomique et de Cosmologie, Universit\'{e} Grenoble-Alpes, CNRS-IN2P3, Grenoble, France\\
$^{80}$ Lawrence Berkeley National Laboratory, Berkeley, California, United States\\
$^{81}$ Lund University Department of Physics, Division of Particle Physics, Lund, Sweden\\
$^{82}$ Moscow Institute for Physics and Technology, Moscow, Russia\\
$^{83}$ Nagasaki Institute of Applied Science, Nagasaki, Japan\\
$^{84}$ Nara Women{'}s University (NWU), Nara, Japan\\
$^{85}$ National and Kapodistrian University of Athens, School of Science, Department of Physics , Athens, Greece\\
$^{86}$ National Centre for Nuclear Research, Warsaw, Poland\\
$^{87}$ National Institute of Science Education and Research, Homi Bhabha National Institute, Jatni, India\\
$^{88}$ National Nuclear Research Center, Baku, Azerbaijan\\
$^{89}$ National Research Centre Kurchatov Institute, Moscow, Russia\\
$^{90}$ Niels Bohr Institute, University of Copenhagen, Copenhagen, Denmark\\
$^{91}$ Nikhef, National institute for subatomic physics, Amsterdam, Netherlands\\
$^{92}$ NRC Kurchatov Institute IHEP, Protvino, Russia\\
$^{93}$ NRC \guillemotleft Kurchatov\guillemotright  Institute - ITEP, Moscow, Russia\\
$^{94}$ NRNU Moscow Engineering Physics Institute, Moscow, Russia\\
$^{95}$ Nuclear Physics Group, STFC Daresbury Laboratory, Daresbury, United Kingdom\\
$^{96}$ Nuclear Physics Institute of the Czech Academy of Sciences, \v{R}e\v{z} u Prahy, Czech Republic\\
$^{97}$ Oak Ridge National Laboratory, Oak Ridge, Tennessee, United States\\
$^{98}$ Ohio State University, Columbus, Ohio, United States\\
$^{99}$ Petersburg Nuclear Physics Institute, Gatchina, Russia\\
$^{100}$ Physics department, Faculty of science, University of Zagreb, Zagreb, Croatia\\
$^{101}$ Physics Department, Panjab University, Chandigarh, India\\
$^{102}$ Physics Department, University of Jammu, Jammu, India\\
$^{103}$ Physics Department, University of Rajasthan, Jaipur, India\\
$^{104}$ Physikalisches Institut, Eberhard-Karls-Universit\"{a}t T\"{u}bingen, T\"{u}bingen, Germany\\
$^{105}$ Physikalisches Institut, Ruprecht-Karls-Universit\"{a}t Heidelberg, Heidelberg, Germany\\
$^{106}$ Physik Department, Technische Universit\"{a}t M\"{u}nchen, Munich, Germany\\
$^{107}$ Politecnico di Bari and Sezione INFN, Bari, Italy\\
$^{108}$ Research Division and ExtreMe Matter Institute EMMI, GSI Helmholtzzentrum f\"ur Schwerionenforschung GmbH, Darmstadt, Germany\\
$^{109}$ Russian Federal Nuclear Center (VNIIEF), Sarov, Russia\\
$^{110}$ Saha Institute of Nuclear Physics, Homi Bhabha National Institute, Kolkata, India\\
$^{111}$ School of Physics and Astronomy, University of Birmingham, Birmingham, United Kingdom\\
$^{112}$ Secci\'{o}n F\'{\i}sica, Departamento de Ciencias, Pontificia Universidad Cat\'{o}lica del Per\'{u}, Lima, Peru\\
$^{113}$ St. Petersburg State University, St. Petersburg, Russia\\
$^{114}$ Stefan Meyer Institut f\"{u}r Subatomare Physik (SMI), Vienna, Austria\\
$^{115}$ SUBATECH, IMT Atlantique, Universit\'{e} de Nantes, CNRS-IN2P3, Nantes, France\\
$^{116}$ Suranaree University of Technology, Nakhon Ratchasima, Thailand\\
$^{117}$ Technical University of Ko\v{s}ice, Ko\v{s}ice, Slovakia\\
$^{118}$ The Henryk Niewodniczanski Institute of Nuclear Physics, Polish Academy of Sciences, Cracow, Poland\\
$^{119}$ The University of Texas at Austin, Austin, Texas, United States\\
$^{120}$ Universidad Aut\'{o}noma de Sinaloa, Culiac\'{a}n, Mexico\\
$^{121}$ Universidade de S\~{a}o Paulo (USP), S\~{a}o Paulo, Brazil\\
$^{122}$ Universidade Estadual de Campinas (UNICAMP), Campinas, Brazil\\
$^{123}$ Universidade Federal do ABC, Santo Andre, Brazil\\
$^{124}$ University of Cape Town, Cape Town, South Africa\\
$^{125}$ University of Houston, Houston, Texas, United States\\
$^{126}$ University of Jyv\"{a}skyl\"{a}, Jyv\"{a}skyl\"{a}, Finland\\
$^{127}$ University of Kansas, Lawrence, Kansas, United States\\
$^{128}$ University of Liverpool, Liverpool, United Kingdom\\
$^{129}$ University of Science and Technology of China, Hefei, China\\
$^{130}$ University of South-Eastern Norway, Tonsberg, Norway\\
$^{131}$ University of Tennessee, Knoxville, Tennessee, United States\\
$^{132}$ University of the Witwatersrand, Johannesburg, South Africa\\
$^{133}$ University of Tokyo, Tokyo, Japan\\
$^{134}$ University of Tsukuba, Tsukuba, Japan\\
$^{135}$ Universit\'{e} Clermont Auvergne, CNRS/IN2P3, LPC, Clermont-Ferrand, France\\
$^{136}$ Universit\'{e} de Lyon, CNRS/IN2P3, Institut de Physique des 2 Infinis de Lyon , Lyon, France\\
$^{137}$ Universit\'{e} de Strasbourg, CNRS, IPHC UMR 7178, F-67000 Strasbourg, France, Strasbourg, France\\
$^{138}$ Universit\'{e} Paris-Saclay Centre d'Etudes de Saclay (CEA), IRFU, D\'{e}partment de Physique Nucl\'{e}aire (DPhN), Saclay, France\\
$^{139}$ Universit\`{a} degli Studi di Foggia, Foggia, Italy\\
$^{140}$ Universit\`{a} di Brescia, Brescia, Italy\\
$^{141}$ Variable Energy Cyclotron Centre, Homi Bhabha National Institute, Kolkata, India\\
$^{142}$ Warsaw University of Technology, Warsaw, Poland\\
$^{143}$ Wayne State University, Detroit, Michigan, United States\\
$^{144}$ Westf\"{a}lische Wilhelms-Universit\"{a}t M\"{u}nster, Institut f\"{u}r Kernphysik, M\"{u}nster, Germany\\
$^{145}$ Wigner Research Centre for Physics, Budapest, Hungary\\
$^{146}$ Yale University, New Haven, Connecticut, United States\\
$^{147}$ Yonsei University, Seoul, Republic of Korea\\

\end{flushleft} 
\end{document}